\documentclass[pdflatex,sn-mathphys-num]{sn-jnl}

\usepackage{float}
\usepackage{multirow}
\usepackage{caption}
\usepackage{subcaption}
\usepackage{graphicx}%
\usepackage{multirow}%
\usepackage{amsmath,amssymb,amsfonts}%
\usepackage{amsthm}%
\usepackage[title]{appendix}%
\usepackage[table]{xcolor}%
\usepackage{textcomp}%
\usepackage{array}
\usepackage{manyfoot}%
\usepackage{booktabs}%
\usepackage{tikz}%
\usetikzlibrary{arrows}
\usepackage{algorithm}%
\usepackage{algorithmicx}%
\usepackage{algpseudocode}%
\usepackage{listings}%
\usepackage{tcolorbox}
\usepackage{colortbl}
\usepackage{float} %
\usepackage{textcomp}
\usepackage{hyperref} %
\usepackage{bookmark} 
\usepackage[british]{babel}
\usepackage [autostyle, english = british]{csquotes}
\MakeOuterQuote{"}
\newcommand{\appref}[1]{Appendix~\ref{#1}}

\theoremstyle{thmstyleone}%
\theoremstyle{thmstyletwo}%

\theoremstyle{thmstylethree}%

\begin{document}

\title[Article Title]{Mapping the landscape of mathematical models for antimicrobial resistance: a scoping review}





\author*[1]{\fnm{Felipe} \sur{Schardong}}\email{felipe.schardong2@gmail.com}

\author[1]{\fnm{Claudio} \sur{Jose Struchiner}}\email{claudio.struchiner@fgv.br}
\equalcont{These authors contributed equally to this work.}

\author[1]{\fnm{Luiz} \sur{Max Carvalho}}\email{luiz.fagundes@fgv.br}
\equalcont{These authors contributed equally to this work.}

\affil[1]{\orgdiv{School of Applied Mathematics}, \orgname{Getulio Vargas Foundation}, \orgaddress{\street{Praia de Botafogo, 190}, \city{Rio de Janeiro}, \postcode{22250-900}, \state{Rio de Janeiro}, \country{Brazil}}}


\abstract{\textbf{Background:} Antimicrobial resistance (AMR) is a serious global public health problem, contributing to an estimated 4.95 million deaths in 2019 and projected to cause up to 10 million annual deaths and a projected US\$100 trillion in cumulative economic losses by 2050. 
Its emergence and spread result from complex biological, ecological, and socioeconomic interactions.
Mathematical modelling has been recognized as a key tool to clarify AMR dynamics. 
However, the dominant literature is fragmented and characterized by notable methodological and contextual limitations. 
This review synthesizes recent mathematical modelling studies on AMR to identify trends, biases, and key research gaps.

\textbf{Methods:} We conducted a scoping review following the PRISMA-ScR. 
Three databases (PubMed, Web of Science, and Scopus) were searched for studies published from 2019 -- 2024 that developed mathematical models about AMR.
After removing duplicates and screening, 36 studies were considered eligible for inclusion.
Data were extracted via a structured form divided into three categories: type and context, construction and correlated parameters, and outputs and validation. 
In each category, the most relevant information for further analysis was extracted.

\textbf{Results:} Our analysis demonstrated a predominance of deterministic models using ordinary differential equations (ODEs), focused on bacteria.
Most studies focused on the human host and one adopting a One Health perspective. 
Conjugation and mutation were the most frequently modelled resistance mechanisms, while transduction, transformation, host immunity, and spatial heterogeneity were rarely included.
Few studies considered economic impact, and a clear geographic imbalance was observed, with most research originating from high-income countries.

\textbf{Conclusion:} Mathematical modelling of AMR is an active field but characterized by low methodological diversity and limited contexts.
There is a need for models that are capable of capturing the complex dynamics among hosts, environments, transmission routes, and intervention.
Deterministic ODE models significantly advance our understanding of AMR dynamics, but future work requires the integration of stochasticity, spatial structure, and ecological interactions to more realistically represent AMR dynamics. 
Incorporating a One Health framework and including economic and social variables will be essential for developing models that not only explain the observed patterns but also guide effective global strategies to mitigate the impact of AMR.

}

\keywords{Mathematical modelling, Drug resistance, Literature review, Epidemiology}



\maketitle

\section{Introduction}
\label{introduction}

Antimicrobial resistance (AMR) occurs when microorganisms develop mechanisms that reduce the efficacy of antimicrobial agents \cite{who2021}.
Recognized by the World Health Organization (WHO) as one of the $10$ greatest global threats of the century and known as the "Silent Pandemic", AMR compromises human and animal health; affects the environment, food, and nutritional security; and hinders economic development and social equality, requiring urgent action and effective management strategies \cite{world2022global}. 
According to projections from $2014$ and $2016$ \cite{review2014, review2016}, approximately $10$ million deaths are expected annually with a cumulative  economic impact of $\$100$ trillion by $2050$. 
More recent studies \cite{murray2022, naghavi2024global}, using data from $2019$ and trends from $1990$ -- $2021$, have confirmed the severity of the problem, particularly in low- and middle-income countries (LMICs), with estimates predicting millions of deaths associated with AMR in the next decades. 
The rise of AMR reduces the efficacy of antimicrobial agents and represents a major threat to modern medicine, as it can result in common infections becoming difficult to treat and compromise procedures such as surgery, chemotherapy, and organ transplantation \cite{puri2025antimicrobial}. 
In addition, AMR causes enormous economic damage to public health systems and the economy in general            \cite{dadgostar2019antimicrobial}, highlighting the need to translate existing knowledge into coordinated and effective containment strategies.

The inappropriate administration of antimicrobials among both humans and animals has long been termed one of the key factors that enhances the development of resistant pathogens and the rise of AMR \cite{machowska2019drivers, sutradhar2014irrational}. 
According to the study by Arepyeva et al. \cite{artigo01}, and under the assumption of their statistical framework, higher antimicrobial consumption was associated with increased resistance, which is consistent with the expectation that stronger selective pressure favours resistance over susceptible subpopulations. 
This type of selection also depends on ecological conditions, the intensity of microbial competition, and the treatment strategies adopted, which can favour or delay the emergence and spread of AMR \cite{artigo08, artigo14}.
The growth of AMR can also be attributed to genetic mechanisms \cite{moo2020mechanisms}, which involve persistent resistance mediated by genes located on chromosomes and plasmids, the latter being particularly advantageous for their ability to perform horizontal gene transfer (HGT), resulting in adaptive advantages in different ecological contexts \cite{artigo11_amr}. 
The results from network-based and spatial structure models have shown that heterogeneous contact between bacteria, as well as the subdivision of the population into demes (subpopulations), can intensify the spread of resistance, enabling the survival of resistant strains even in hostile selective environments \cite{artigo13_amr, artigo31}.

Another determining factor in maintaining resistance is related to nosocomial infections. 
Hospital environments, with strong selective pressure and a high presence of antimicrobial agents, favour the endemicity of resistant strains,  complicating epidemiological control \cite{goldmann1996strategies, sydnor2011hospital}.
Mathematical models developed for these environments show that, although antimicrobials are effective against sensitive strains, inadequate treatment of patients carrying resistant strains can establish resistance in the hospital population \cite{artigo12_amr}.
Innovative strategies, such as therapeutic combinations, are increasingly being explored. 
The use of phages combined with antibiotics, for example, has shown good results and proven to be a promising alternative for combating AMR \cite{artigo41}. 
Similarly, adaptive treatments that explore competition for resources between resistant and susceptible strains can restore the effectiveness of traditional antimicrobial agents \cite{artigo14}.

From a health policy perspective, it is widely accepted that reducing the risks associated with AMR depends on an integrated approach based on the One Health concept.
This strategy enables convergence involving human, animal, and environmental health and has been proposed as key to guiding public policy and practices in governance \cite{artigo09_amr, artigo11}. 
Consequently, effective policy formulation and resource allocation depend on mathematical models that effectively capture these multisectoral interactions and ecological feedbacks.
Awareness campaigns are also very powerful tools for combating AMR and have had a positive impact on reducing inappropriate antimicrobial prescription and adjusting population behaviour regarding antimicrobial use, with the impact depending on the structure and duration of the campaign \cite{artigo12}. 
Another very effective method is drug risk management programs, which are recognized as key strategies for optimizing the use of antimicrobials, reducing selective pressure, and helping to prevent the spread of AMR \cite{artigo13}.

In this context, mathematical modelling has been established as an essential tool for understanding the dynamics of AMR. 
Models may offer hypotheses regarding the spread and transmission of pathogens, the development and evolution of AMR, and the effects of interventions at various scopes.
For example, compartmental models have been used to examine how antibiotic consumption and transmission dynamics influence resistance at the population level \cite{niewiadomska2019population} and to inform stewardship and infection control strategies in healthcare settings \cite{durazzi2023modelling}.
Evolutionary and within-host models have provided insights into the mechanisms of mutation, selection, and horizontal gene transfer leading to multidrug resistance \cite{roberts2021combining}.
Agent-based and network models have also been employed to capture heterogeneity in patient contact structures and hospital transmission dynamics \cite{donker2012hospital}. 
This significant methodological variety provides varied insights into AMR, while simultaneously posing challenges in the comparison, integration, and transformation of the findings to applicable recommendations for decision-makers and policy-makers.

Despite significant advances in the mathematical modelling of AMR, research in this field remains concentrated on specific approaches.
Most existing studies rely on deterministic models that are based on ODEs, whose mathematical simplicity and low computational cost facilitate initial analyses. 
However, this simplicity often limits the ability to capture stochasticity, individual heterogeneity, and spatial structure, factors essential for a realistic understanding of AMR dynamics and evolution. 
Moreover, the diversity of assumptions, data scarcity, and heterogeneous epidemiological contexts hinder the development of a unified perspective on AMR.
In this context, there remains a lack of systematic synthesis regarding how these models are constructed, which assumptions and limitations recur across them, and how such constraints affect progress in the field. 
To address these gaps, this study conducts a scoping review of mathematical modelling applied to AMR, aiming to identify and characterize the types of models used and the pathogens and antimicrobial treatments modelled in the studies. 
By providing a structured overview, this review seeks to offer a conceptual framework to support the development of more realistic and integrative models capable of addressing critical gaps in current AMR mathematical modelling research.

\subsection{Background on mathematical models}

Mathematical models of AMR provide a structured way to represent how sensitive and resistant populations emerge, change, and respond to selective pressures across biological scales, including antimicrobial action, microbial evolution, and ecological or epidemiological interactions \cite{holmes2016}.
Although these models differ in their levels of biological and mathematical complexity, the majority have a common goal: to capture the mechanisms that drive the emergence and spread of resistance \cite{lipsitch2002}.
We review the technical aspects of AMR mathematical modelling to help readers interpret the results of this review.
In what follows, we provide a concise, structured summary of conceptual components that commonly characterize within-host formulations, since these elements recur across many of the approaches discussed in the text and in the papers included in this review.

A first essential component is the pharmacokinetic/pharmacodynamic (PK/PD) relationship that links the within-host time evolution of the drug concentration, typically denoted by the function \(A(t)\) (PK), with the effect of that concentration on the bacteria (PD), represented by a function \(E(A)\).
In practice, \(E(A)\) is often modelled via Emax/Hill-type functions \cite{regoes2004}, which translate the intensity of bacteriostatic or bactericidal effects as a function of antimicrobial exposure.
The PK component encompasses absorption, distribution, metabolism, and excretion (ADME) processes, whereas the PD component describes how drug exposure affects the survival of sensitive and resistant strains, and together these two elements define the selective pressures imposed by treatment and shape the emergence of resistance, even though many models simplify this block because detailed PK/PD data are often unavailable \cite{drusano2004}.

A second common component in within-host models concerns microbial dynamics, which are typically represented by equations describing the evolution of sensitive (\(S\)) and resistant (\(R\)) populations over time. These formulations combine terms for growth, competition, generation of resistance, antimicrobial-induced mortality, and, in some models, host immune action. Bacterial growth is often modelled with logistic-type terms such as

\begin{equation}
\alpha_i\,X\left(1-\frac{S+R}{K}\right), \qquad i,X \in \{S,R\},
\end{equation}
while the treatment effect is incorporated via exposure-dependent mortality terms, for example \(E_i(A)\,X\) with \(i, X\in\{S,R\}\) \cite{lipsitch1997, otto2011}. 
Resistance emergence typically appears as either (i) a mutation flux \(S\to R\) modelled by a term \(\nu S\), or (ii) horizontal gene transfer (HGT), which is frequently represented by conjugation and written as a mass-action term \(\lambda S R\) \cite{zurwiesch2011}. 
Other HGT mechanisms, such as transduction or transformation, are less often included because they are harder to parameterise. 
Many models also incorporate fitness costs, implemented as a reduced growth rate for \(R\) (i.e., \(\alpha_R<\alpha_S\)). More general extensions may introduce multiple resistant classes (\(R_1,R_2,\dots\)) to represent resistance to different drugs (multidrug resistance, MDR) \cite{almutairy2024extensively}.

Mathematical models can be implemented via different formalisms. 
Deterministic models, often expressed through systems of ordinary differential equations (ODEs), describe how the mean behavior of the system evolves over time and are the most common owing to their analytical tractability and lower computational cost \cite{otto2011}.
In contrast, stochastic models explicitly incorporate random events, such as rare mutations or sporadic horizontal gene transfer (HGT) events, which can substantially influence the emergence of resistance, especially when pathogen populations are small \cite{allen2010introduction}. Stochastic simulations, such as those generated by the Gillespie algorithm, allow the exploration of variability and uncertainty in resistance trajectories \cite{gillespie1977}. 
Other approaches include partial differential equation (PDE) models to capture spatial heterogeneity or agent-based models (ABMs) to simulate interactions at the individual level \cite{macal2005}.

\begin{figure}[H]
\centering
\makebox[\linewidth]{%
\includegraphics[scale=0.34]{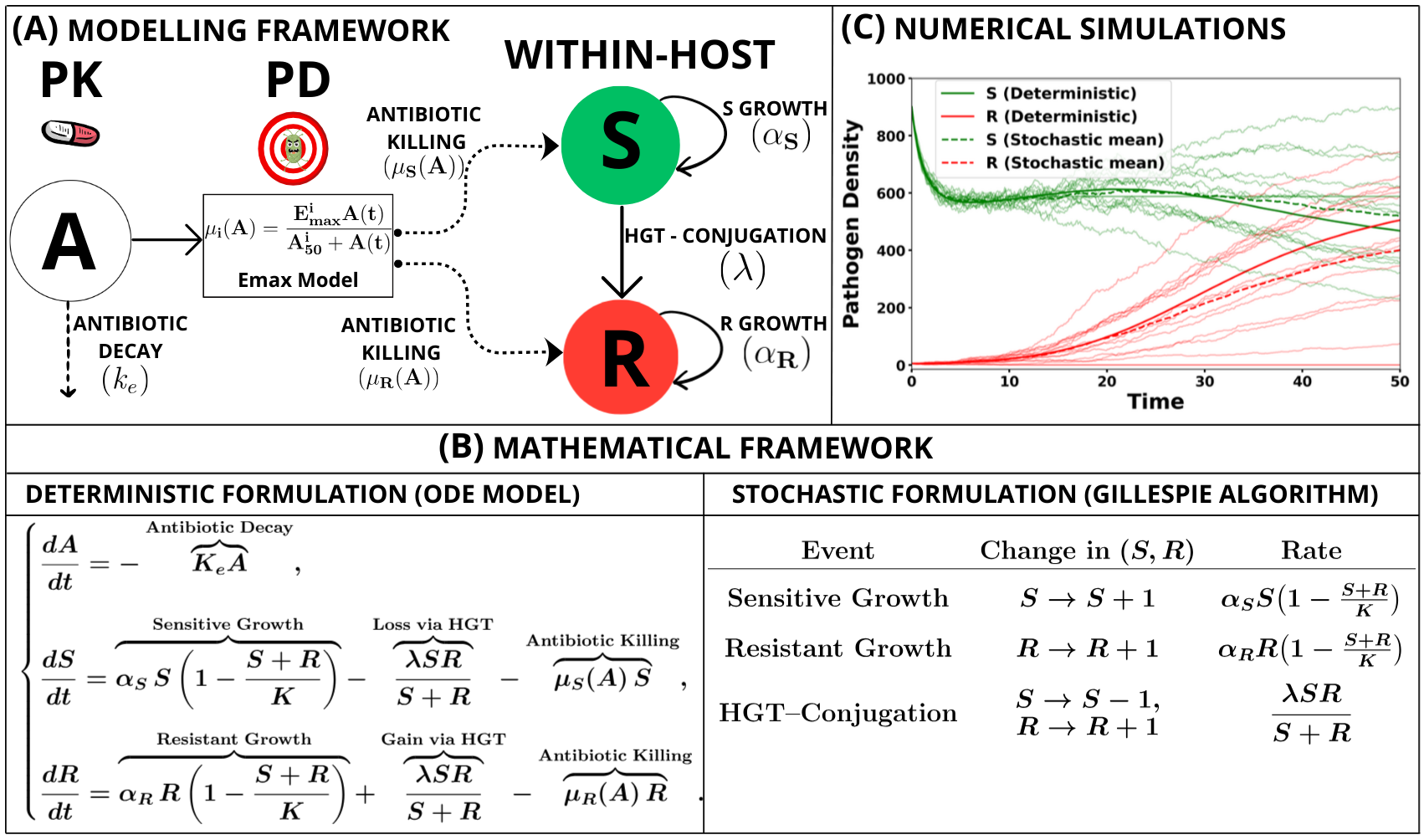}
}
\caption{\textbf{Comprehensive modelling framework for AMR, combining conceptual design, mathematical formulations, and numerical simulations.} (A) Schematic representation of the full modelling framework, integrating PK/PD and within-host bacterial dynamics. The PK module governs the antibiotic concentration $A(t)$, which feeds into the PD killing functions $\mu_i(A)$, while sensitive ($S$) and resistant ($R$) bacterial populations interact through growth, antibiotic killing, and horizontal gene transfer (HGT). (B) Mathematical formulations of the model. Left: deterministic ordinary differential equations describing PK, PD, bacterial growth, and HGT. Right: stochastic version based on the Gillespie algorithm, listing each event, state transition, and corresponding rate. (C) Numerical simulations comparing deterministic trajectories of sensitive and resistant populations with multiple stochastic realizations and their mean behaviour.}
\label{fig:background}
\end{figure}

Figure \ref{fig:background} summarizes these elements in a compact visual representation and shows a simplified version of an AMR model taken from Techitnutsarut \& Chamchod (2021) \cite{artigo12_amr_mtm}, which was included in our review.
Panel (A) illustrates how PK/PD dynamics and within-host pathogen dynamics are typically integrated in a single model. Panel (B) gives an example of how such systems can be formalized in deterministic (ODE) and stochastic (Gillespie) frameworks. 
Panel (C) shows how these formulations translate into numerical simulations, comparing mean trajectories and stochastic variability for sensitive and resistant populations.
Since the goal of this exposition is purely illustrative, we adopted initial conditions and parameter values that differ from those used in the original paper; for the numerical simulations shown in panel (C) we set: \(k_{e} = 0.1\,\mathrm{h}^{-1}\), \(\alpha_{S} = 1\), \(\alpha_{R} = 0.8\), \(\lambda = 0.01\), \(E_{max}^{S} = 0.5\,\mathrm{h}^{-1}\), \(E_{max}^{R} = 0.2\,\mathrm{h}^{-1}\), \(K = 1000\), \(A(0)=10\ \mu\mathrm{g}/\mathrm{ml}\), \(S(0)=900\) and \(R(0)=5\).

The conceptual elements described here (PK/PD relationships, within-host dynamics, and deterministic and stochastic formulations) represent only a subset of the components that constitute mathematical models of AMR. 
To systematically map how these and other modelling aspects have been applied in recent literature, and identify trends, methodological biases, and gaps, we conducted a scoping review following the PRISMA-ScR guidelines.
Section \ref{methods} details the internal protocol, eligibility criteria, search strategy (databases consulted, $2019$ -- $2024$), screening procedures, and structured data-extraction form used for the included articles, which organized the information into the three categories employed in the analysis.
These methodological procedures support the presentation of results in Section \ref{results}.

\section{Methods}
\label{methods}
   The study adhered to the guidelines and checklist of the Preferred Reporting Items for Systematic Reviews and Meta-Analyses extension for scoping reviews (PRISMA-ScR) \cite{tricco2018prisma}. 
   All items from the PRISMA-ScR checklist were considered in this review, except for the critical appraisal of individual sources.
   This decision reflects the principal goal of our study, which is to map the literature on the mathematical modelling of AMR and to identify research gaps to inform future research directions, rather than to assess the methodological quality of each included study.

   \subsection{Protocol and registration}
   \label{protocol_and_registration}
   No specific protocol was developed or registered prior to conducting this review. 
Although the creation and registration of a protocol are widely recommended to enhance transparency, reduce duplication of efforts, and improve reporting quality \cite{sacks1987meta}, this study did not have a formally registered protocol publicly available or accessible upon request.
To ensure methodological rigor, however, an internal protocol was developed and adhered to by the research team, serving as a framework to guide the planning, conduct, and reporting of the review. 
The methodological approach followed the Arksey and O’Malley framework for scoping reviews \cite{arksey2005scoping}, which was subsequently refined by Levac et al. \cite{levac2010} and by the Joanna Briggs Institute (JBI) guidance \cite{peters2020scoping}. 
Future reviews building on the present findings may benefit from formal protocol registration on platforms such as PROSPERO or OSF.
   
   \subsection{Eligibility criteria}
   \label{eligibility_criteria}
   
We followed the PCC (Population, Concept, Context) framework recommended by the JBI \cite{peters2020scoping} to guide the objective, research question, and inclusion criteria. 
An overview of the literature on the mathematical modelling of AMR is the scope of this review, with the search and selection of articles prioritizing studies that develop or apply mathematical models to describe the emergence, spread, and control of drug-resistant microbial populations. 
Based on the PCC framework, the following inclusion rules were defined:

\begin{enumerate}
\item \textbf{Population (P):} studies applying AMR modelling to humans, animals, in vitro systems, combinations of these, or general contexts; the pathogens considered included bacteria, viruses, fungi, and parasites.

\item \textbf{Concept (C):} studies employing dynamic mathematical models to investigate AMR and its mechanisms, focusing on aspects such as model type and context, construction and correlated parameters, and outputs and validation.

\item \textbf{Context (C):} studies conducted in intrahost, community, hospital/ICU, laboratory, or general settings, indexed in Scopus, Web of Science, or PubMed, published in English between January 1, 2019, and December 31, 2024.
\end{enumerate}

The exclusion criteria included studies related to AMR without a formal mathematical structure or lacking a dynamic modelling component, articles that did not develop original models (e.g., purely descriptive works or statistical analyses without population dynamics), book chapters, narrative or systematic reviews without a new model proposal, conference abstracts, non-peer-reviewed material, papers published in languages other than English, and studies whose full text was unavailable. 
The period ($2019$ -- $2024$) was selected to capture recent methodological advances, reflect the continuous growth of the literature in the area, and allow for an updated analysis of the main trends in the mathematical modelling of AMR.

   \subsection{Information sources}
   \label{information surces}
   We searched Scopus, PubMed, and Web of Science for relevant literature, selecting these databases for their extensive coverage of mathematical modelling studies and rigorous indexing standards. 
The search was conducted between 1 January 2019 and 31 December 2024 and used a single Boolean string, adapted for each database, combining terms related to AMR (e.g., antimicrobial, antibiotic, antiviral, antifungal, antiparasitic resist*), resistance-transfer mechanisms (e.g., mutation, horizontal gene transfer, plasmid, conjugation, transduction, phage), and modelling approaches (e.g., mathematical model, compartmental model, agent-based model, stochastic model). 
All the search results were exported in CSV format, and duplicates were automatically removed via Python script available in our GitHub repository (see \appref{code_and_data}). 
Two reviewers independently screened all titles and abstracts for eligibility, resolving discrepancies by consensus, with a third reviewer consulted when necessary. 
The reference lists of the included studies were also examined manually to identify additional relevant papers. 
Grey literature (e.g., preprints, dissertations, and technical reports) was excluded to maintain a focus on peer-reviewed research.
   
   \subsection{Screening process}
   \label{screening_process}
   
\begin{figure}[H] 
\centering
\begin{tikzpicture}[>=latex, font={\sf \small}]

\tikzstyle{bluerect} = [rectangle, rounded corners, minimum width=1.5cm, minimum height=0.75cm, text centered, draw=black, fill=cyan!60!gray!45!white, rotate=90, font={\sffamily \large}]

\tikzstyle{roundedrect} = [rectangle, rounded corners, minimum width=12cm, minimum height=1cm, text centered, draw=black, font={\sffamily \Large}]

\tikzstyle{textrect} = [rectangle, minimum width=5.25cm, text width=5.24cm, minimum height=1cm, draw=black, font={\sffamily \normalsize}]

\node (top1) at (0, 10.5cm) [draw, roundedrect, fill=yellow!80!red!70]
  {Identification of studies via database};

\node (r1blue) at (-6.75cm, 8.0cm) [draw, bluerect, minimum width=3cm]{Identification};

\node (r1left) at (-3.25cm, 8.0cm) [draw, textrect, minimum height=2.5cm]
  {Records identified from:
     \begin{itemize}
     \item Web of Science ($n=1112$)
     \item PubMed ($n=1029$)
     \item Scopus ($n=1658$)
     \end{itemize}   
  };

\node (r1right) at (3.25, 8.0cm) [draw, textrect, minimum height=2.5cm]
  {Records removed before screening: 
    \begin{itemize}
    \item Duplicates ($n=1322$)
    \item Ineligible ($n=12$)
    \end{itemize} 
  };

\node (r2blue) at (-6.75cm, 2.2cm) [draw, bluerect, minimum width=6.8cm]
  {Screening};

\node (r2left) at (-3.25cm, 4.7cm) [draw, textrect, minimum height=1.5cm]
  {Records screened ($n=2465$)};

\node (r2right) at (3.25, 4.7cm) [draw, textrect, minimum height=1.5cm]
  {Records excluded ($n= 2394$)
  \begin{itemize}
     \item title/abstract not relevant
     \item not meeting eligibility criteria
   \end{itemize}};

\node (r4left) at (-3.25cm, 2.2cm) [textrect, minimum height=1.5cm]
  {Assessed for eligibility ($n= 71$)};

\node (r6left) at (-3.25cm, -0.3cm) [textrect, minimum height=1.5cm]
  {Eligible studies ($n= 22$)};  

\node (r4right) at (3.25cm, 2.2cm) [textrect, minimum height=.5cm]
  {Reports excluded: ($n= 49$)
    \begin{itemize}
    \item Adherence to the actual topic of AMR modelling (e.g. studies only considering HGT in general were excluded)
    \item Availability of the full-text
    \end{itemize}     
  };

\node (r6right) at (3.25cm, -1.6cm) [textrect, minimum height=1.5cm]
  {Reference tracking: ($n=14$)};

\draw[->] (r4left.east) -- (r4right.west);

\coordinate (midpoint_rt) at (-3.25cm, -1.6cm);  
\draw[->] (r6right.west) -- (midpoint_rt);

\node (r5blue) at (-6.75cm, -3.0cm) [draw, bluerect, minimum width=2cm]
  {Included};

\node (r5left) at (-3.25cm, -3.0cm) [draw, textrect, minimum height=1.5cm]
  {Included in final review ($n= 36$)};

\draw[thick, ->] (r1left) -- (r1right);
\draw[thick, ->] (r1left) -- (r2left);
\draw[thick, ->] (r2left) -- (r2right);
\draw[thick, ->] (r2left) -- (r4left);
\draw[thick, ->] (r4left) -- (r6left);
\draw[thick, ->] (r6left) -- (r5left);

\end{tikzpicture}
\caption{PRISMA flow diagram of the study selection process.}
\label{prisma_flow_diagram}
\end{figure}
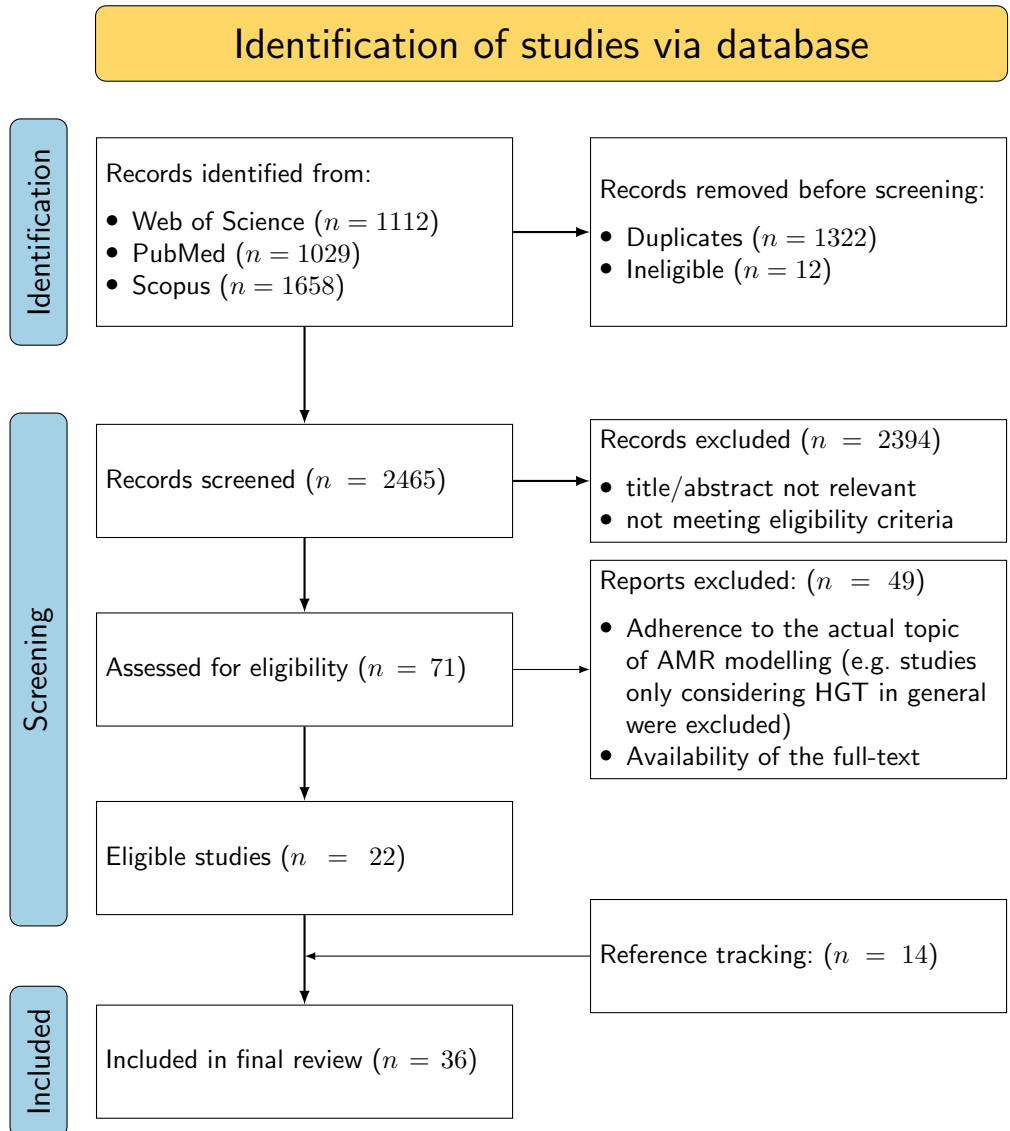

The complete screening process is shown in Figure \ref{prisma_flow_diagram}. The searches were conducted in three databases on April 11, 2025, using the following terms:

\begin{itemize}
    \item \textbf{Web of Science search:} ALL=((antimicrobial OR antibiotic OR antiviral OR antifungal OR antiparasitic)resist* AND (mechanism* OR mutation* OR "horizontal gene transfer" OR plasmid OR transformation OR conjugation OR transduction OR phage) AND (math* OR dynamic* OR stochastic OR deterministic OR compartmental OR "agent-based") model*), $1112$ results.
    
     \vspace{0.5cm}
     
    \item \textbf{PubMed search:} (antimicrobial OR antibiotic OR antiviral OR antifungal OR antiparasitic)resist* AND (mechanism* OR mutation* OR "horizontal gene transfer" OR plasmid OR transformation OR conjugation OR transduction OR phage) AND ((math* OR dynamic* OR stochastic OR deterministic OR compartmental OR "agent-based") model*), $1029$ results. 
    
     \vspace{0.5cm}
     
    \item \textbf{Scopus search:} TITLE-ABS-KEY((antimicrobial OR antibiotic OR antiviral OR antifungal OR antiparasitic) AND (resist* OR tolerance OR persistence) AND (mechanism* OR mutation* OR "horizontal gene transfer" OR plasmid OR transformation OR conjugation OR transduction OR phage) AND (math* OR dynamic* OR stochastic OR deterministic OR compartmental OR "agent-based") AND model*) AND PUBYEAR $>$ $2018$ AND PUBYEAR $<$ $2025$, $1658$ results.
\end{itemize}

Our study selection process, involving the identification, screening, and inclusion of relevant papers, is summarized in Fig. \ref{prisma_flow_diagram}. 
Initially, we retrieved $3.799$ articles, of which some $1.322$ were discarded for duplication and some $12$ were discarded because they were ineligible. The duplicate removal process was performed in Python, based on the DOI and title.
All the articles exported from the three databases are available in the \texttt{all\_data} folder in the GitHub repository (see \appref{code_and_data}).
One author read the title and abstract of the remaining $2.465$ articles and, based on the research objectives, selected $71$ articles using the pre-established eligibility criteria in the Eligibility Criteria Subsection \ref{eligibility_criteria}.

These 71 studies were then independently evaluated by a second author, who reviewed the full texts.
This step resulted in the exclusion of 49 studies.
The primary reasons were non-adherence to the core topic of AMR mechanistic modelling (e.g., models of general HGT not applied to AMR) and unavailability of the full text. 
Finally, to ensure comprehensiveness, we performed backwards reference tracking on the $22$ included studies, which identified an additional $14$ eligible studies, resulting in a final total of $36$ studies for qualitative analysis in this review.

   \subsection{Information extracted from the included papers}
   \label{information_extracted}
   We developed a data charting form to record the variables relevant to the scope of the study. 
Following the recommendations of Levac et al. \cite{levac2010}, the form was iteratively updated as the team conducted the review and analysis of the included studies $(n = 36)$. 

The three categories analysed contain relevant information for understanding how mathematical models are developed and applied to the study of AMR. 
Specifically, the data in Table \ref{table_data_charting_form} encompass (i) the epidemiological and biological context of the model (Type and Context); (ii) the structural and mechanistic aspects involved in model construction (Construction and Correlated Parameters); and (iii) the analytical strategies and evaluated results (Outputs and Validation).
These elements provide an overview of how modelling studies address the complexity of AMR systems.

The first category, Type and Context, addresses the research setting in which the model is defined.
It includes the mathematical formulation (e.g., deterministic or stochastic), the model structure (e.g., ODEs, PDEs), and the host type under consideration (e.g., human, animal, or in vitro). 
It also integrates the broader epidemiological setting (e.g., hospital, community, intra-host) and the pathogens addressed, whether at the category level (e.g., bacteria, virus) or at the level of specific organisms (e.g., \textit{E. coli}, \textit{S. aureus}).
These characteristics define the validity domain of the model and establish the biological and epidemiological context for interpreting its results.

The second category, Construction and Correlated Parameters, encompasses the internal mechanisms and biological assumptions implemented in the models.
It records how resistance is represented (mutation, horizontal gene transfer), whether pharmacokinetic/pharmacodynamic (PK/PD) processes are explicitly included, and whether spatial dispersion is modelled. 
Additional elements include host immune-response dynamics that act on pathogen clearance, fitness costs for resistant strains, the possibility of multiple resistances, and explicit drug concentration thresholds (e.g., MICs). 
Together, these characteristics determine the structural complexity of the models and influence their ability to capture biological realism.

The third category, Results and Validation, focuses on the strategies employed to analyse and test the models. It includes whether treatment strategies are simulated (e.g., monotherapy, polytherapy), whether sensitivity analyses are performed to assess parameter influence, and whether external validation is attempted. 
The outputs considered may include predictions of resistance evolution, evaluation of economic impacts, or the identification of practical applications for policy or clinical decision-making.
These aspects provide insight into the robustness, applicability, and translational potential of the modelling frameworks.

Together, these three categories, summarized in Table \ref{table_data_charting_form}, provide a framework that captures the epidemiological and biological context, the structural mechanisms of the models, and the analytical results of the studies, including whether the model incorporated an economic component (e.g., cost or cost-benefit considerations) when applicable. 
The structured data presented in this table formed the basis for the next steps of the thematic and quantitative analysis, allowing the identification of trends, gaps, and clusters in the existing literature on mathematical modelling of AMR covered in this scoping review.
The detailed list of aspects extracted from each study can be found in \appref{aspects_of_each_study}.
The file \texttt{questionnaire\_answers.csv}, containing all the extracted elements from the 36 studies included in this review, is available in our GitHub repository (\appref{code_and_data}).

\begin{table}[htbp] 
\centering
\caption{Summary of categories and attributes extracted from the selected models}
\begin{tabular}{ |p{2.9cm}|p{3.9cm}|p{4.9cm}| >{\raggedright\arraybackslash}p{3.0cm}|} 
\hline
\rowcolor{gray!25}\rule{0pt}{20pt}{\centering\textbf{Model Category}\par} & 
{\centering\textbf{Recorded Information}\par} & {\centering\textbf{Signification}\par} & 
{\centering\textbf{Possible Values}\par} \\
\hline

\rule{0pt}{15pt} \multirow{22}{7em}{\centering\textbf{Type and Context}} & Model type & Model type used in the study & {\footnotesize Deterministic OR Stochastic OR Both} \\ 

\rule{0pt}{15pt} & Mathematical structure & Mathematical structure of the model & {\footnotesize ODEs OR PDEs OR SDEs OR IDEs }\\ 

\rule{0pt}{15pt} & Host & Type of host represented in the model & {\footnotesize Human OR Animal OR In-Vitro OR Human and Animal OR General }\\ 

\rule{0pt}{15pt} & System context &  System setting represented in the model & {\footnotesize Within-Host OR Community OR Hospital OR Laboratory OR General} \\ 

\rule{0pt}{15pt} & Pathogen category & Pathogen category represented in the model & {\footnotesize Virus OR Bacteria OR Both} \\ 

\rule{0pt}{15pt} & Specific pathogen & Specific pathogen(s) included in the model & {\footnotesize \textit{P. aeruginosa} OR \textit{E. coli} OR \textit{S. aureus} OR \textit{A. baumanni} OR \textit{S. pneumoniae} OR \textit{C. difficie} OR \textit{K. pneumoniae} OR HIV OR Combination} \\ \hline

\rule{0pt}{15pt} \multirow{26}{7em}{\centering\textbf{Construction and Correlated Parameters}} & Acquired resistance & Mechanism of resistance acquisition modelled & {\footnotesize Mutations OR Mutation and HGT–Conjugation OR HGT–Conjugation OR HGT–Conjugation and Transformation OR HGT–transformation OR N/A}
\\

\rule{0pt}{15pt} & PK/PD parameters & Type of Pharmacokinetic/Pharmacodynamic (PK/PD) component included in the model & {\footnotesize PK/PD OR PK OR PD OR N/A} \\

\rule{0pt}{15pt} & Spatial dispersion & Spatial consideration in the model & {\footnotesize Yes OR No} \\

\rule{0pt}{15pt} & Fitness cost & Fitness cost for the growth of resistant strains & {\footnotesize Yes OR No} \\

\rule{0pt}{15pt} & Immune response & Presence of host immune response in the model & {\footnotesize Yes OR No} \\

\rule{0pt}{15pt} & Multiple resistance & Consideration of multiple AMR resistance in the model & {\footnotesize Yes OR No} \\

\rule{0pt}{15pt} & Drug concentration & Threshold drug concentration for antimicrobial efficacy (e.g., MIC) & {\footnotesize Yes OR No} \\
\hline

\rule{0pt}{15pt} \multirow{15}{7em}{\centering\textbf{Outputs and Validation}} & Treatment strategy & Antimicrobial treatment strategy in the model & {\footnotesize Monotherapy OR Polytherapy OR Both} \\

\rule{0pt}{15pt} & Sensitivity analysis & Impact of parameters on model outputs & {\footnotesize Yes OR No} \\

\rule{0pt}{15pt} & External validation & Compare predictions with other models/studies & {\footnotesize Yes OR No} \\

\rule{0pt}{15pt} & Evolution of resistance & Changes in resistance over time within the model & {\footnotesize Yes OR No} \\

\rule{0pt}{15pt} & Economic impact & Economic impact related to AMR & {\footnotesize Yes OR No} \\

\rule{0pt}{15pt} & Practical applications & The model results can be used in practical applications & {\footnotesize Yes OR No} \\
\hline
\end{tabular}
\label{table_data_charting_form}
\end{table}

\newpage

\section{Results}
\label{results}

\begin{figure}[H]
  \centering
  \makebox[\linewidth]{%
    \includegraphics[scale=0.27]{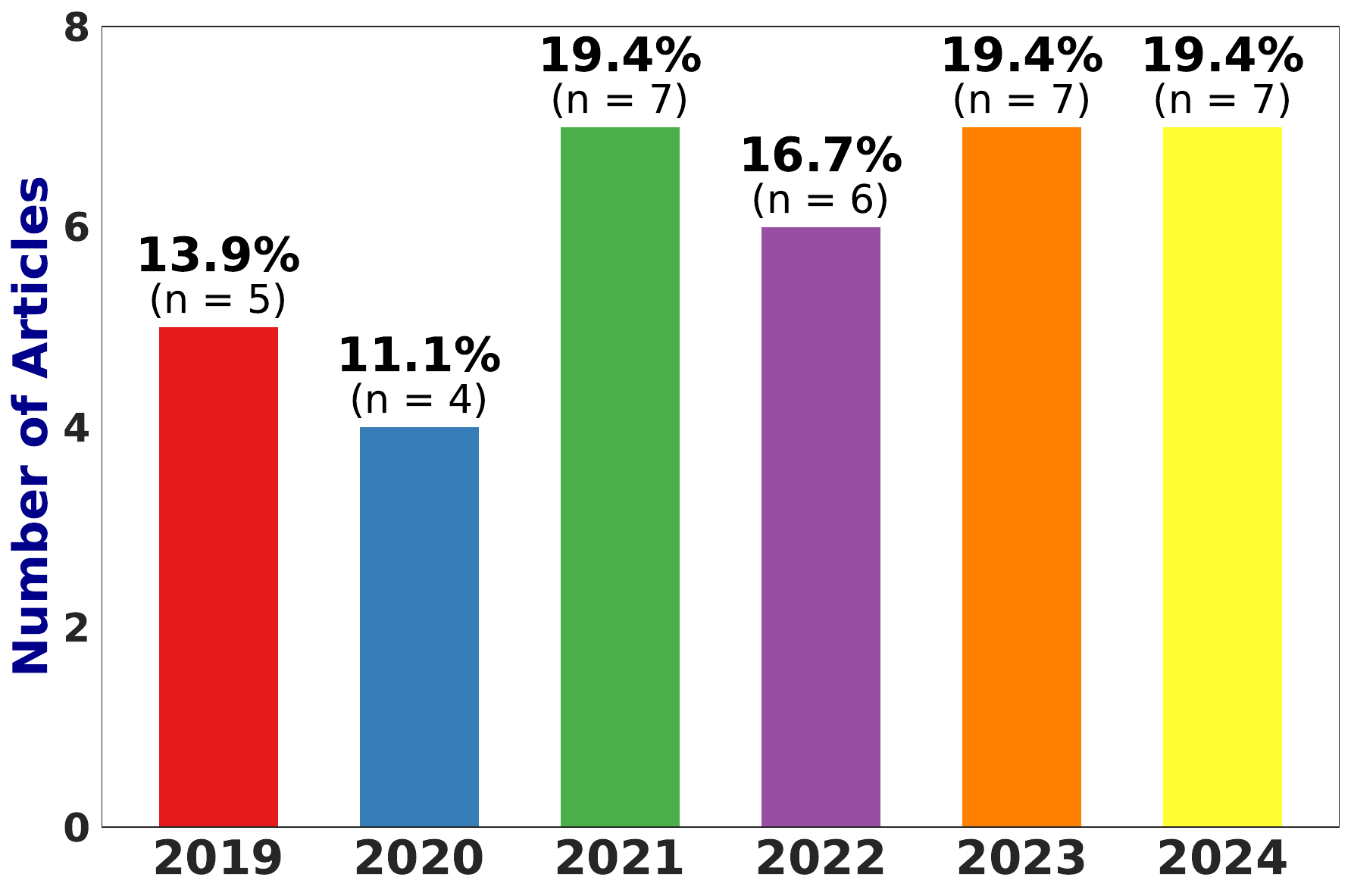}
  }
  \caption{\textbf{Temporal distribution of the 36 studies included in the scoping review (2019–2024).} Illustrating a consistent research output in the mathematical modelling of AMR over the period.}
  \label{fig:papers_by_year}
\end{figure}

A total of $36$ studies were included in this scoping review \cite{artigo01_amr_mtm,artigo02_amr_mtm,artigo03_amr_mtm,artigo04_amr_mtm,artigo05_amr_mtm, artigo06_amr_mtm,artigo07_amr_mtm,artigo08_amr_mtm,artigo09_amr_mtm,artigo10_amr_mtm,
artigo11_amr_mtm,artigo12_amr_mtm,artigo13_amr_mtm,artigo14_amr_mtm,artigo15_amr_mtm,
artigo16_amr_mtm,artigo17_amr_mtm,artigo18_amr_mtm,artigo19_amr_mtm,artigo20_amr_mtm,
artigo21_amr_mtm,artigo22_amr_mtm,artigo23_amr_mtm,artigo24_amr_mtm,artigo25_amr_mtm,
artigo26_amr_mtm,artigo27_amr_mtm,artigo28_amr_mtm,artigo29_amr_mtm,artigo30_amr_mtm,
artigo31_amr_mtm,artigo32_amr_mtm,artigo33_amr_mtm,artigo34_amr_mtm,artigo35_amr_mtm,
artigo36_amr_mtm}. 
The period from $2019$ -- $2024$ was selected to capture the most recent developments in the mathematical modelling of AMR, reflecting the increasing interest and methodological diversification observed in recent years. 
The temporal distribution of the publications is shown in Fig. \ref{fig:papers_by_year}. Five articles $(13.5\%)$ were published in $2019$, four $(10.8\%)$ in $2020$, and eight $(21.6\%)$ in $2021$. Six studies $(16.2\%)$ were published in $2022$, while the most recent years, $2023$ and $2024$, each contributed seven articles $(18.9\%)$. 
This distribution highlights a consistent research effort in modelling AMR dynamics during the period considered.

\begin{figure}[H]
  \centering
  \makebox[\linewidth]{%
    \includegraphics[scale=0.5]{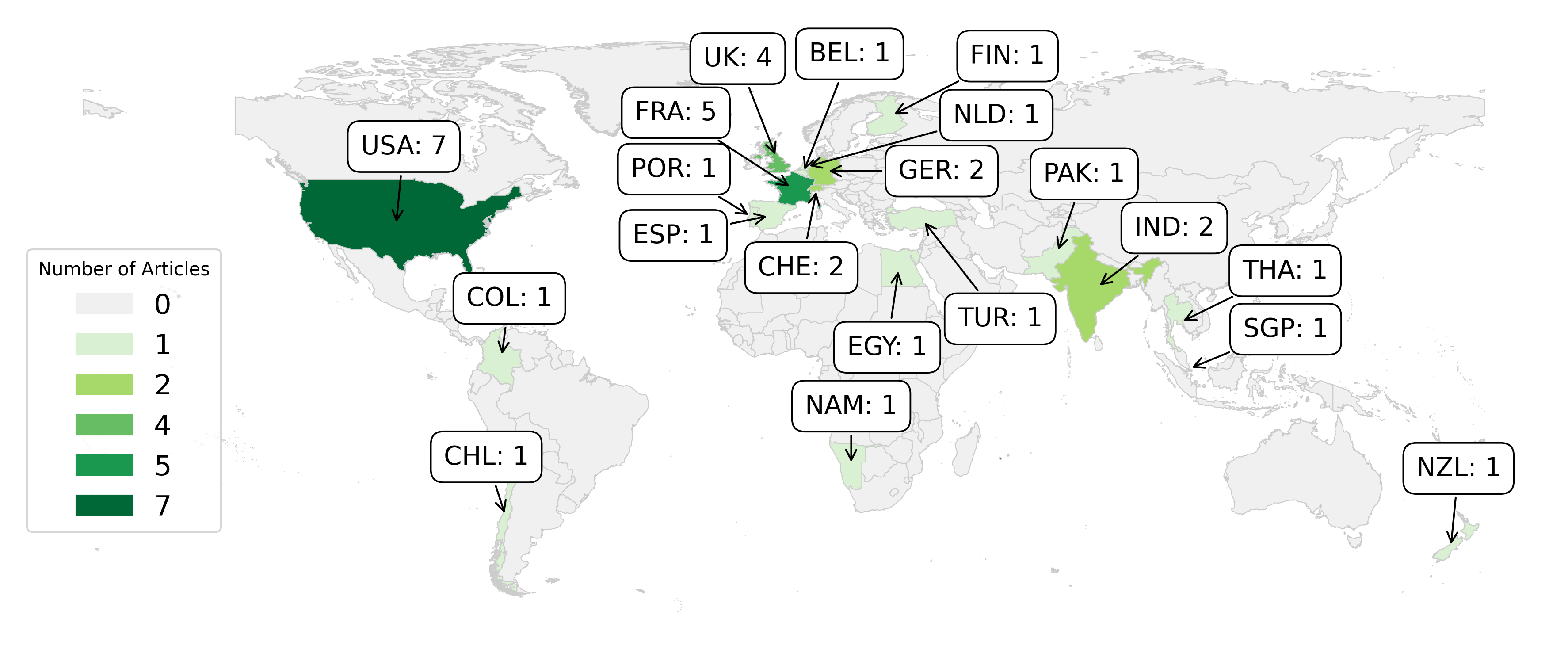}
  }
  \caption{\textbf{National affiliation of the first author's institution for articles included in the scoping review.} The figure highlights the predominance of studies from high-income countries (HICs) and the underrepresentation of contributions from the Global South.}
  \label{fig:national_affiliation_author}
\end{figure}

The $36$ included studies were affiliated with institutions from $20$ different countries.
The most represented were the United States (seven studies), France (five), and the United Kingdom (four). 
Several other high-income countries contributed fewer studies, including Germany and Switzerland (two each). Portugal, Belgium, the Netherlands, New Zealand, Singapore, Spain, Finland, and Turkey (one each).
Countries from the Global South, such as India (two studies), Egypt, Pakistan, Colombia, Chile, Namibia, and Thailand, were underrepresented, with only one study each. This distribution highlights a clear geographical imbalance in research on mathematical modelling of AMR, with a predominance of high-income countries (HICs) (see Fig. \ref{fig:national_affiliation_author}).

\subsection{Model type and context}

The reviewed studies were first categorized on the basis of their methodological approach and the epidemiological context they aimed to represent. In particular, we examined the mathematical formulation of the models (e.g., deterministic, stochastic, etc), as well as the epidemiological scales represented.
This analysis highlights a clear predominance of certain model types and mathematical structures, choices that are often influenced by the research question, the extent and precision of available data, and the necessity to reconcile analytical tractability with computational feasibility. 
Moreover, the distribution of host organisms, system contexts, and pathogen types offers valuable insights into the main priorities and limitations guiding current research.
These categories provide a comprehensive view of how modelling efforts have been framed in the AMR field, showing prevailing practices and underexplored directions.

\begin{figure}[H]
  \centering
  \begin{subfigure}{0.48\textwidth}
     \includegraphics[scale=0.26]{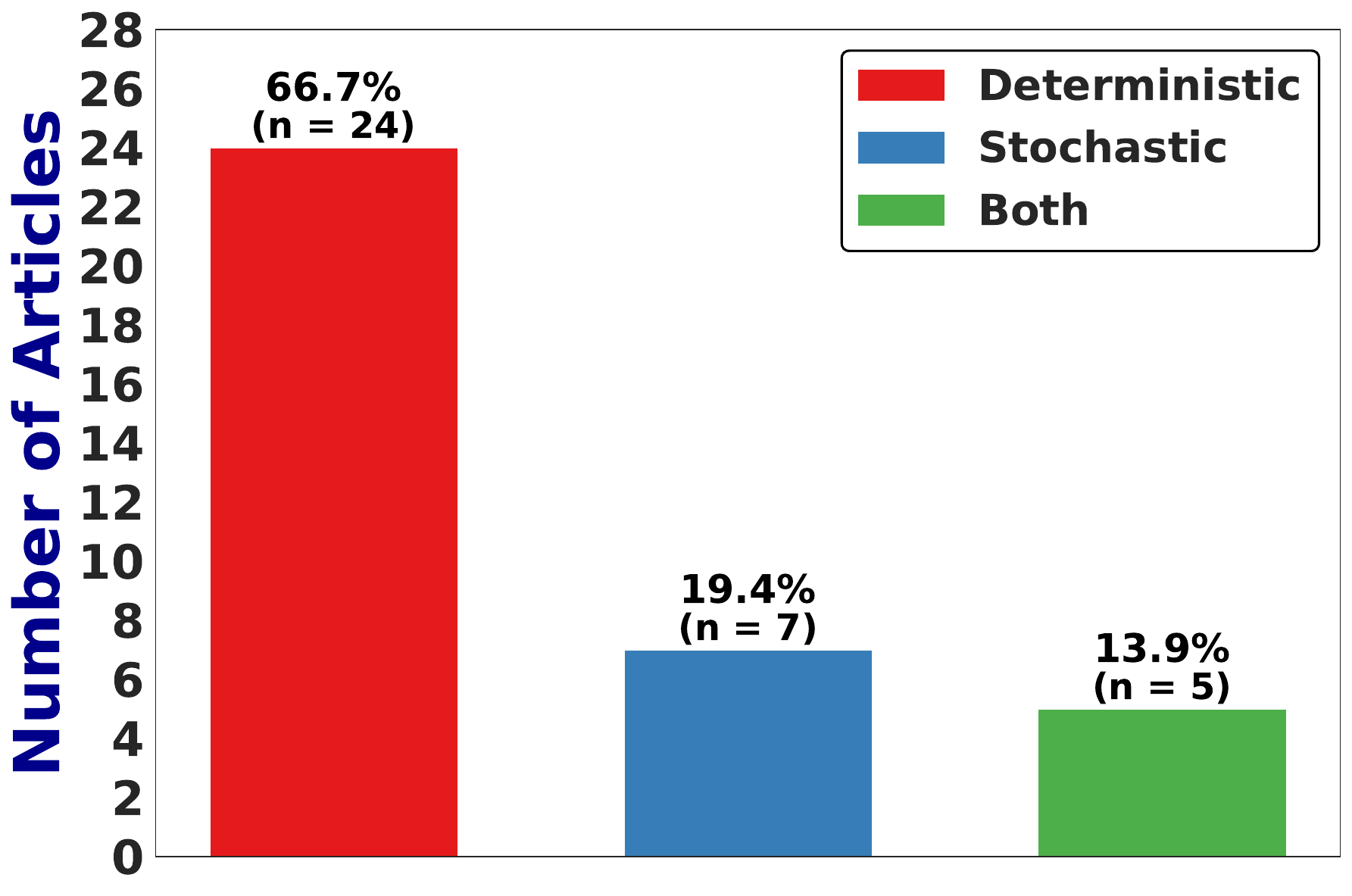}
     \caption{Model type}
     \label{fig:model_type}
  \end{subfigure}
  \quad
  \begin{subfigure}{0.48\textwidth}
     \includegraphics[scale=0.26]{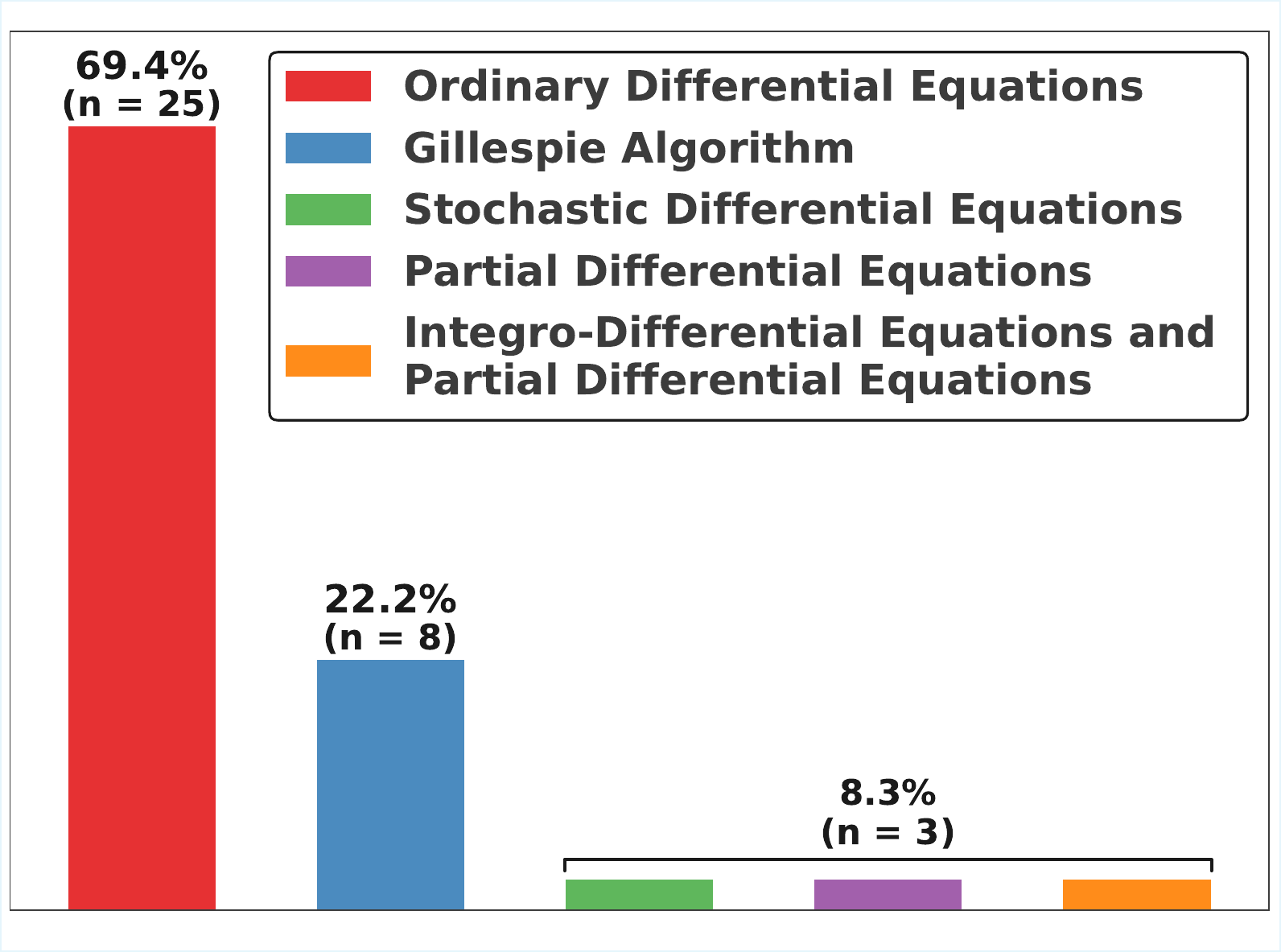}
     \caption{Mathematical structure}
     \label{fig:mathematical_structure}
  \end{subfigure}

  \caption{\textbf{Model type and mathematical structure used in the reviewed studies.} (a) Distribution of model types, highlighting the predominance of deterministic models over stochastic and hybrid approaches. (b) Mathematical structures adopted, with most studies employing ordinary differential equations (ODEs), followed by stochastic differential equations (SDEs), partial differential equations (PDEs), and integro-differential equations (IDEs).}
  \label{fig:model_type_and_mathematical_structure}
\end{figure}

Figure \ref{fig:model_type_and_mathematical_structure} illustrates the results of the $36$ studies related to the type of model and the mathematical structure used. Figure \ref{fig:model_type} presents the distribution of model types adopted in the reviewed studies, highlighting that twenty-four articles $(66.7\%)$ utilized deterministic models \cite{artigo02_amr_mtm,artigo03_amr_mtm,artigo04_amr_mtm,artigo06_amr_mtm,artigo07_amr_mtm,artigo08_amr_mtm,artigo09_amr_mtm,artigo11_amr_mtm,artigo12_amr_mtm,artigo13_amr_mtm,artigo14_amr_mtm,artigo17_amr_mtm,artigo19_amr_mtm,artigo21_amr_mtm,artigo25_amr_mtm,artigo26_amr_mtm,artigo27_amr_mtm,artigo29_amr_mtm,artigo30_amr_mtm,artigo31_amr_mtm,artigo32_amr_mtm,artigo33_amr_mtm,artigo34_amr_mtm,artigo36_amr_mtm}, seven studies $(19.4\%)$ analysed stochastic models \cite{artigo01_amr_mtm,artigo16_amr_mtm,artigo18_amr_mtm,artigo22_amr_mtm,artigo23_amr_mtm,artigo24_amr_mtm,artigo35_amr_mtm} and five $(13.9\%)$ conducted analyses and comparisons between both models \cite{artigo05_amr_mtm,artigo10_amr_mtm,artigo15_amr_mtm,artigo20_amr_mtm,artigo28_amr_mtm}. 
Now, for the mathematical structure (Figure \ref{fig:mathematical_structure}) we find $25$ studies $(69.4\%)$ that use ordinary differential equations (ODEs) \cite{artigo02_amr_mtm,artigo03_amr_mtm,artigo04_amr_mtm,artigo05_amr_mtm,artigo06_amr_mtm,artigo07_amr_mtm,artigo08_amr_mtm,artigo09_amr_mtm,artigo11_amr_mtm,artigo12_amr_mtm,artigo13_amr_mtm,artigo14_amr_mtm,artigo15_amr_mtm,artigo16_amr_mtm,artigo17_amr_mtm,artigo19_amr_mtm,artigo25_amr_mtm,artigo26_amr_mtm,artigo27_amr_mtm,artigo29_amr_mtm,artigo31_amr_mtm,artigo32_amr_mtm,artigo33_amr_mtm,artigo34_amr_mtm,artigo36_amr_mtm}, eight articles $(22.2\%)$ adopted the Gillespie algorithm \cite{artigo10_amr_mtm,artigo18_amr_mtm,artigo20_amr_mtm, artigo22_amr_mtm,artigo23_amr_mtm,artigo24_amr_mtm, artigo28_amr_mtm, artigo35_amr_mtm}, one study used stochastic differential equations (SDEs) \cite{artigo01_amr_mtm}, one used partial differential equations (PDEs) \cite{artigo21_amr_mtm}
 and one utilized integro-differential equations (IDEs) and PDEs \cite{artigo30_amr_mtm}, each representing one $2.8\%$ of the studies.
 Among the articles that used Gillespie's algorithm, \cite{artigo20_amr_mtm,artigo23_amr_mtm,artigo24_amr_mtm,artigo28_amr_mtm,artigo35_amr_mtm} applied Gillespie's exact algorithm. Articles \cite{artigo10_amr_mtm,artigo18_amr_mtm} used approximations, such as the $\tau$-leap algorithm or its approximation, which is based on the binomial distribution \cite{artigo22_amr_mtm}.
 Although article \cite{artigo35_amr_mtm} uses Gillespie's algorithm to simulate its birth–death model, it employs linear noise approximation (LNA), a technique that approximates discrete stochastic dynamics via an SDE for the purpose of moment discovery and analysis.
 The article \cite{artigo01_amr_mtm} developed a primarily continuous approach, modelling the dynamics through an SDE and using the Euler–Maruyama method for numerical simulation.

The predominance of deterministic models in the reviewed literature, especially those based on ODEs, is due to their relatively simple implementation, lower computational cost, and analytical tractability \cite{brauer2017mathematical}. 
These models allow researchers to investigate the population-level dynamics of AMR and evaluate the effects of interventions on the basis of simplified representations of biological and epidemiological processes, which can be valuable when detailed data are scarce \cite{lehtinen2017evolution,niewiadomska2019population}. 
In contrast, stochastic or high-dimensional models often require more detailed parametrisation and computational resources, making them less feasible in settings with limited or aggregate data \cite{allen2010introduction,dureau2013capturing}. 
Moreover, software tools commonly used in mathematical epidemiology, such as MATLAB, R, and Python, are well-established numerical solvers for ODEs, facilitating their adoption. 
For example, MATLAB provides the ODE suite (ode45, ode15s, among others) \cite{shampine1997matlab}, R includes packages such as deSolve for deterministic model implementation \cite{soetaert2010solving}, and Python offers the SciPy.integrate module (\texttt{solve\_ivp}, odeint) for flexible ODE solving \cite{virtanen2020scipy}.
This methodological accessibility reinforces the preference for ODE-based approaches, even in complex problems such as AMR. Nevertheless, this trend may also reflect limitations in available data, as more complex models (e.g., stochastic or spatial) typically require finer-grained information that is often unavailable or difficult to obtain in AMR surveillance contexts \cite{laxminarayan2013antibiotic, grundmann2011framework}.

Although EDO-based models offer ease of analysis, the strong dependence on these deterministic structures still imposes an important methodological limitation for public health applications.
The emergence and spread of AMR, particularly the emergence of resistance (rare events) or rapid changes in small populations (e.g., in hospitals), are essentially stochastic processes.
The scarcity of models that capture stochasticity (only $19.4\%$) reduces the predictive power of analyses and compromises their usefulness in complex and realistic epidemiological contexts. 
This demonstrates the need for current modelling practices to capture real-world dynamics and effectively respond to practical challenges.

\begin{figure}[H]
  \centering
  \begin{subfigure}{0.48\textwidth}
     \includegraphics[scale=0.26]{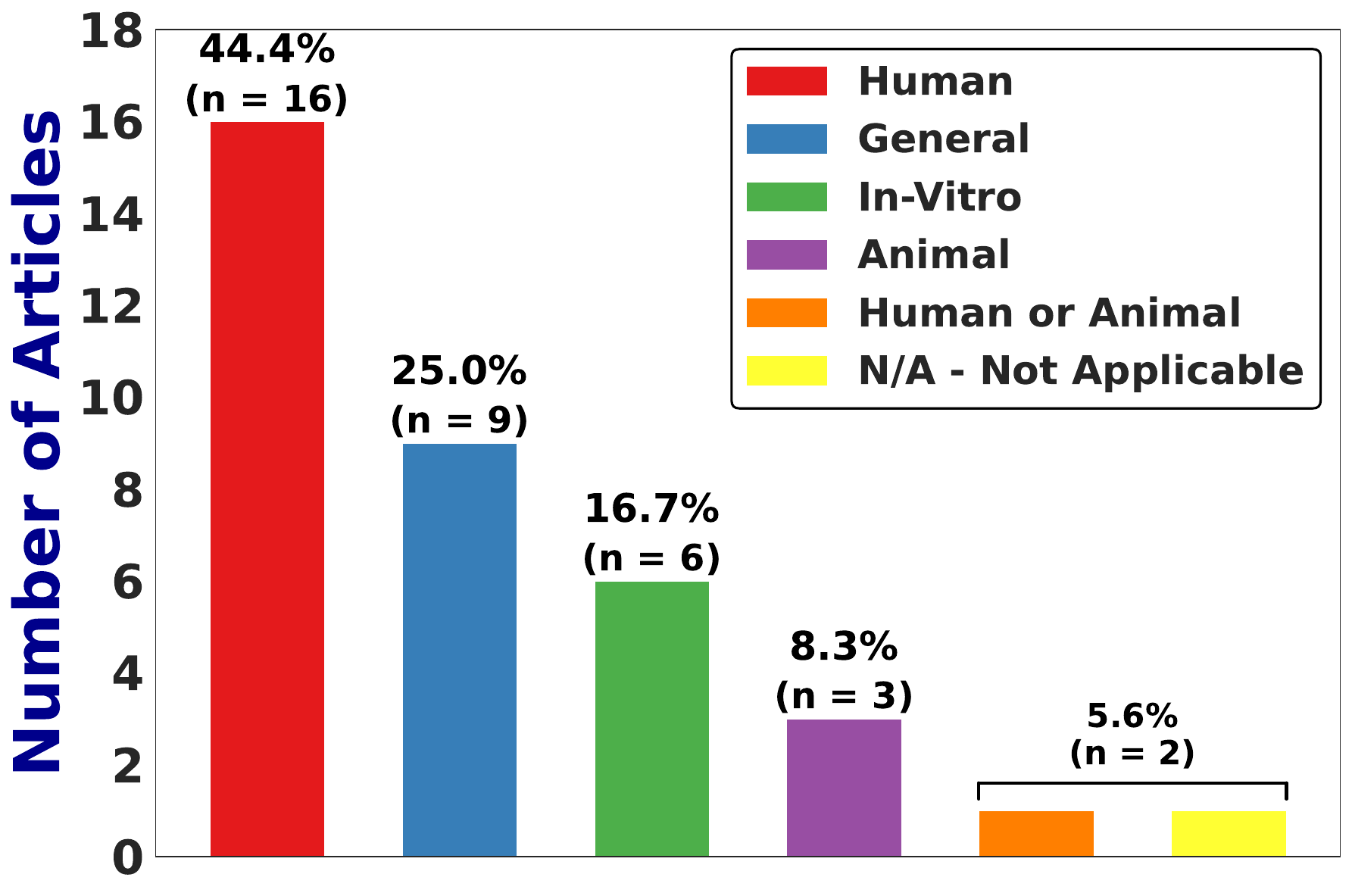}
     \caption{Host}
     \label{fig:host}
  \end{subfigure}
  \quad
  \begin{subfigure}{0.48\textwidth}
     \includegraphics[scale=0.26]{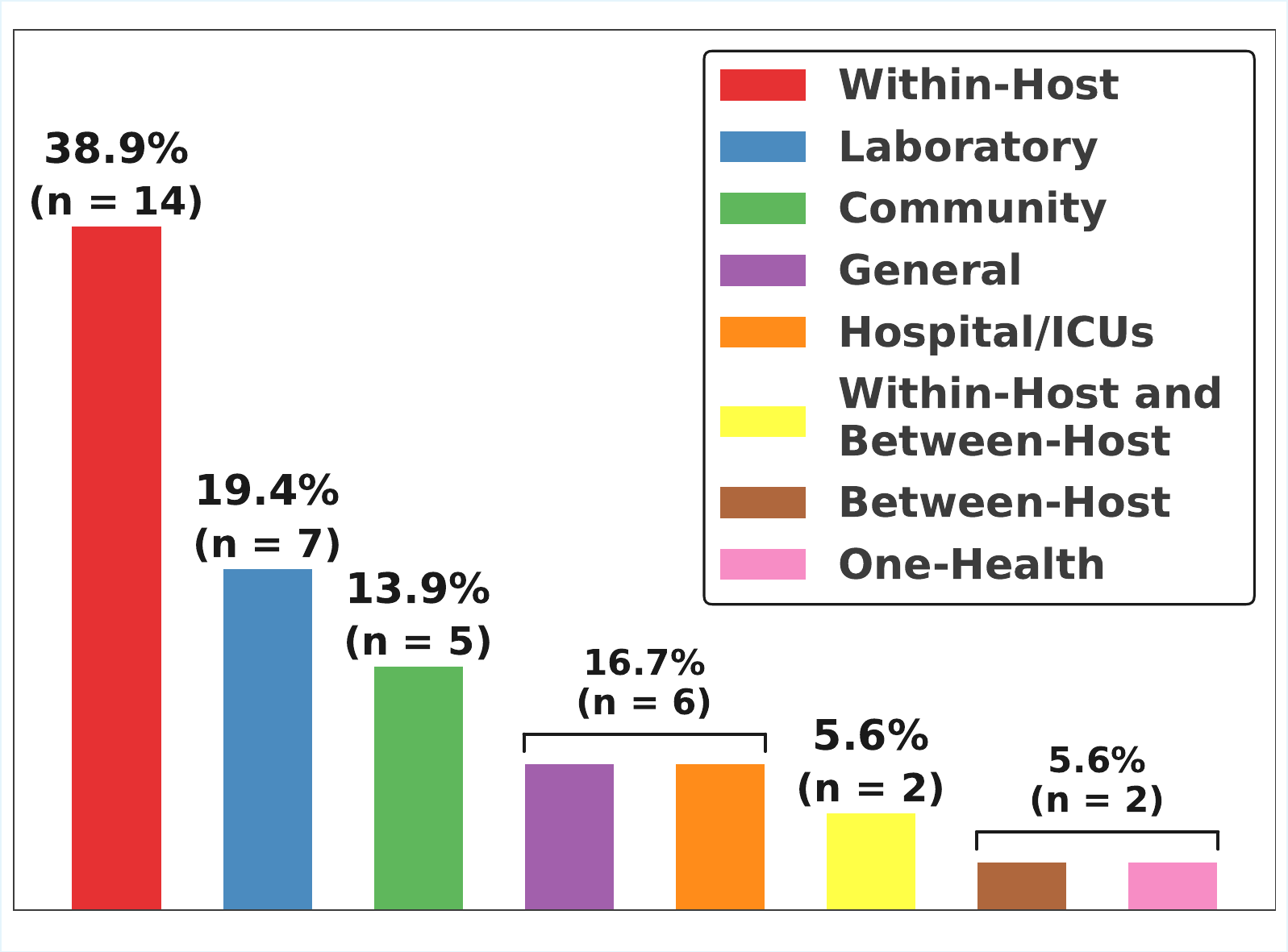}
     \caption{System context}
     \label{fig:system_context}
  \end{subfigure}

  \caption{\textbf{Host organisms and epidemiological contexts represented in the selected studies.} (a) Distribution of host types modelled, including human, animal, in vitro, general, human or animal, and only animal types. (b) System-level contexts explored, such as within-host, laboratory, community, hospital, or unspecified settings.}
  \label{fig:host_and_system_context}
\end{figure}

Figure \ref{fig:host_and_system_context} provides an overview of the host organisms modelled and the system-level settings explored in the selected studies. As shown in Figure \ref{fig:host}, human host were the most common, used in nineteen studies $(44.4\%)$ \cite{artigo04_amr_mtm,artigo05_amr_mtm,artigo08_amr_mtm,artigo09_amr_mtm,artigo10_amr_mtm,artigo12_amr_mtm,artigo15_amr_mtm,artigo18_amr_mtm,artigo23_amr_mtm,artigo27_amr_mtm,artigo29_amr_mtm,artigo30_amr_mtm,artigo31_amr_mtm,artigo32_amr_mtm,artigo33_amr_mtm,artigo36_amr_mtm}, and the general host appeared in nine articles $(25\%)$ \cite{artigo01_amr_mtm,artigo02_amr_mtm,artigo07_amr_mtm,artigo11_amr_mtm,artigo13_amr_mtm,artigo16_amr_mtm,artigo21_amr_mtm,artigo24_amr_mtm,artigo28_amr_mtm}, in vitro appeared in six studies $(16.7\%)$ \cite{artigo06_amr_mtm,artigo19_amr_mtm,artigo20_amr_mtm,artigo22_amr_mtm,artigo34_amr_mtm,artigo35_amr_mtm}, three articles $(8.3\%)$ \cite{artigo14_amr_mtm,artigo25_amr_mtm,artigo26_amr_mtm} used models that can be applied to animals; one study $(2.8\%)$ \cite{artigo03_amr_mtm} included both humans and animals, and one article \cite{artigo17_amr_mtm} did not apply to any host.  
With respect to the modelled epidemiological context (Figure \ref{fig:system_context}), more than one-third of the studies $(14/36)$ focused on within-host dynamics \cite{artigo01_amr_mtm,artigo02_amr_mtm,artigo04_amr_mtm,artigo05_amr_mtm,artigo07_amr_mtm,artigo11_amr_mtm,artigo12_amr_mtm,artigo13_amr_mtm,artigo14_amr_mtm,artigo15_amr_mtm,artigo16_amr_mtm,artigo23_amr_mtm,artigo26_amr_mtm,artigo33_amr_mtm}. 
Laboratory settings were represented in seven studies $(19.4\%)$ \cite{artigo06_amr_mtm,artigo19_amr_mtm,artigo20_amr_mtm,artigo22_amr_mtm,artigo25_amr_mtm,artigo34_amr_mtm,artigo35_amr_mtm}. Five studies $(3.9\%)$ examined community-level contexts \cite{artigo08_amr_mtm,artigo09_amr_mtm,artigo17_amr_mtm,artigo18_amr_mtm,artigo27_amr_mtm}. 
Three studies $(8.3\%)$ addressed general modelling frameworks \cite{artigo21_amr_mtm,artigo24_amr_mtm,artigo28_amr_mtm}. The other three studies $(8.3\%)$ focused on hospital/ICU environments \cite{artigo31_amr_mtm,artigo32_amr_mtm,artigo36_amr_mtm}. Two studies $(5.6\%)$ combined within- and between-host scales \cite{artigo10_amr_mtm,artigo30_amr_mtm}.  
One study $(2.8\%)$ investigated between-host dynamics \cite{artigo03_amr_mtm} and one study focused on One-Health contexts \cite{artigo29_amr_mtm}. 

These results highlight the strong predominance of models focused on human hosts, emphasizing the clinical priority of AMR in public health contexts \cite{review2016}.
Additionally, the low representation of animal hosts and the absence of models that include several sectors (One Health) suggest an important gap, with recent reviews highlighting that multi-scalar models are rare, given their great potential to capture crucial characteristics that drive AMR \cite{booton2021one}.
In general, these studies reveal a disconnected scenario, reflecting both limited multidisciplinarity and the absence of data across human, animal, and environmental sectors \cite{morel2020}. 
Future mathematical models should use multi-scalar and integrated approaches, adopting the One Health perspective, and explicitly connect human hosts, animals, and environments to effectively capture the complex ecology of AMR \cite{white2019critical,tetteh2020survey}.

\begin{figure}[H]
  \centering
  \begin{subfigure}{0.48\textwidth}
     \includegraphics[scale=0.26]{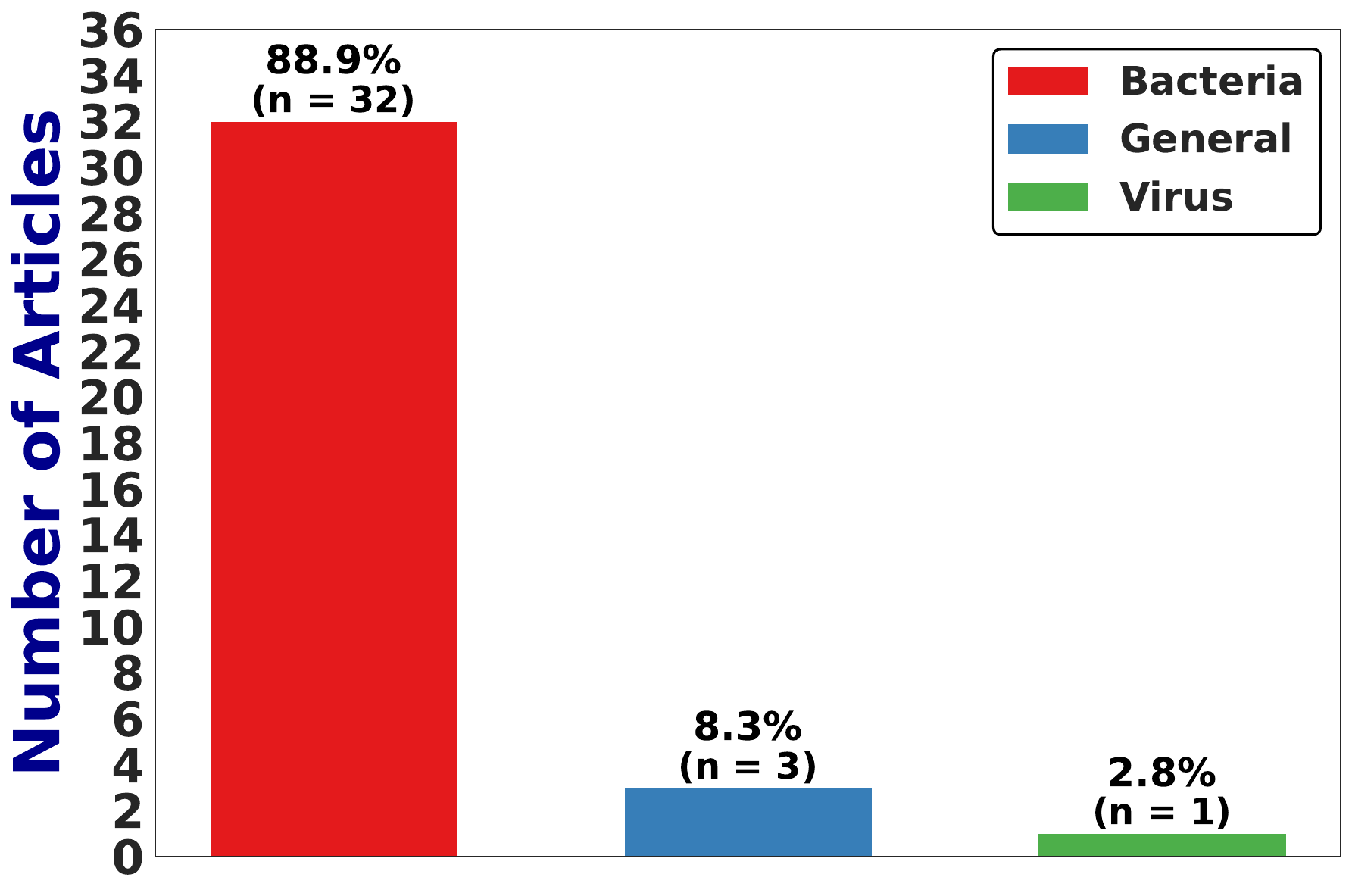}
     \caption{Pathogen category}
     \label{fig:pathogen}
  \end{subfigure}
  \quad
  \begin{subfigure}{0.48\textwidth}
     \includegraphics[scale=0.26]{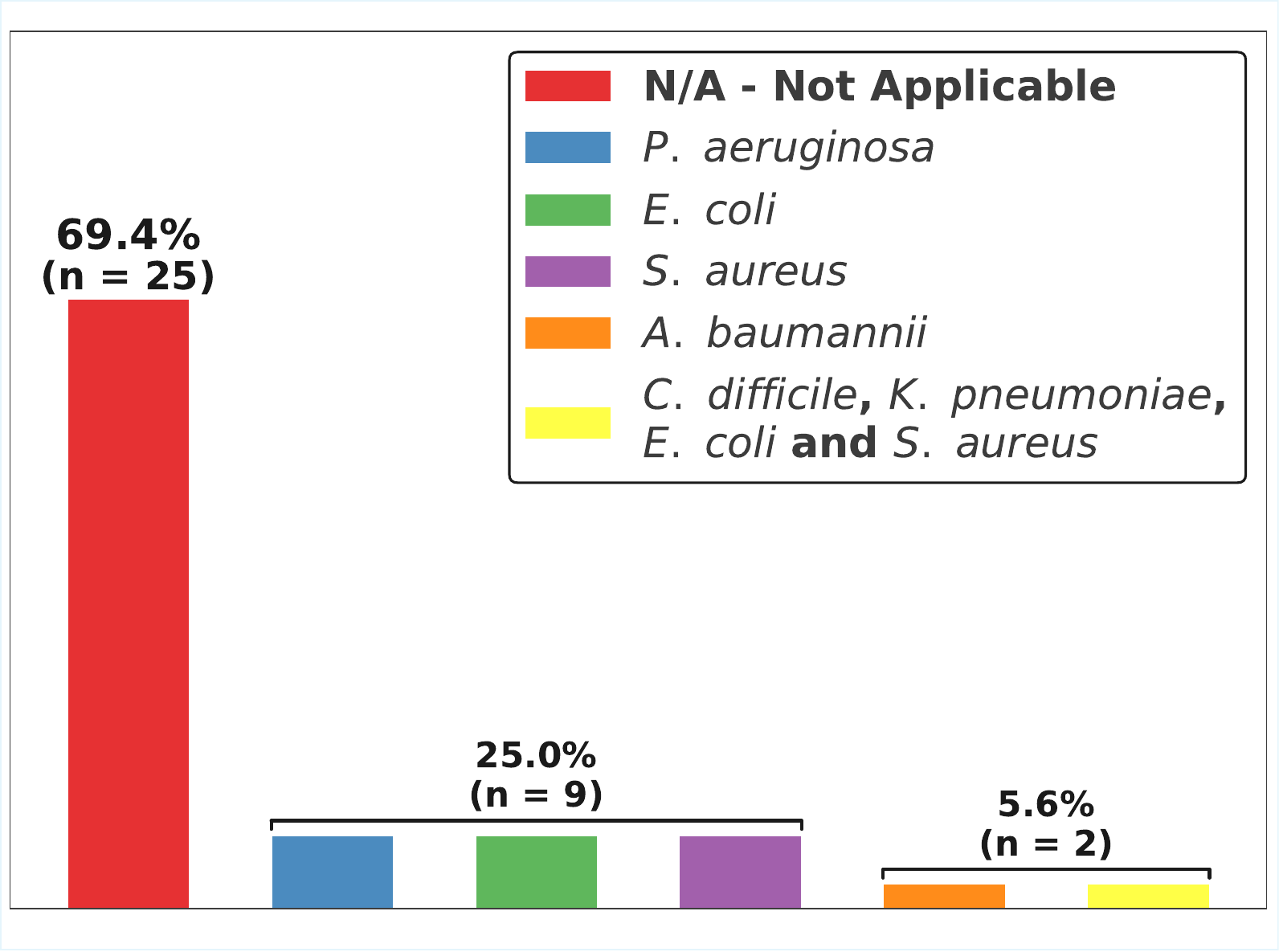}
     \caption{Specific pathogen}
     \label{fig:specific_pathogen}
  \end{subfigure}

  \caption{\textbf{Pathogen categories and specific pathogens modelled in the selected studies.} (a) Distribution of general pathogen types, highlighting the predominance of bacterial models. (b) Specific pathogens addressed in the models, including common AMR-related bacteria and one viral pathogen (HIV).}
  \label{fig:pathogen_and_specific_pathogen}
\end{figure}

Figure \ref{fig:pathogen_and_specific_pathogen} summarizes the types of pathogens addressed in the selected studies.
As shown in Figure \ref{fig:pathogen}, the vast majority $(32/36)$ of the models focused on bacteria \cite{artigo01_amr_mtm,artigo02_amr_mtm,artigo03_amr_mtm,artigo04_amr_mtm,artigo06_amr_mtm,artigo07_amr_mtm,artigo08_amr_mtm,artigo09_amr_mtm,artigo11_amr_mtm,artigo12_amr_mtm,artigo13_amr_mtm,artigo14_amr_mtm,artigo15_amr_mtm,artigo16_amr_mtm,artigo17_amr_mtm,artigo18_amr_mtm,artigo19_amr_mtm,artigo20_amr_mtm,artigo21_amr_mtm,artigo22_amr_mtm,artigo23_amr_mtm,artigo25_amr_mtm,artigo26_amr_mtm,artigo27_amr_mtm,artigo29_amr_mtm,artigo30_amr_mtm,artigo31_amr_mtm,artigo32_amr_mtm,artigo33_amr_mtm,artigo34_amr_mtm,artigo35_amr_mtm,artigo36_amr_mtm}, whereas only one study $(2.8\%)$ addressed a viral infection \cite{artigo05_amr_mtm}, and another three $(8.3\%)$ adopted a general pathogen category without specifying the type \cite{artigo10_amr_mtm,artigo24_amr_mtm,artigo28_amr_mtm}. In terms of specificity (Figure \ref{fig:specific_pathogen}), a significant portion of the studies $(25/36)$ did not explicitly model a specific pathogen (N/A) \cite{artigo01_amr_mtm,artigo02_amr_mtm,artigo03_amr_mtm,artigo04_amr_mtm,artigo05_amr_mtm,artigo06_amr_mtm,artigo08_amr_mtm,artigo09_amr_mtm,artigo10_amr_mtm,artigo11_amr_mtm,artigo12_amr_mtm,artigo13_amr_mtm,artigo14_amr_mtm,artigo15_amr_mtm,artigo16_amr_mtm,artigo17_amr_mtm,artigo18_amr_mtm,artigo21_amr_mtm,artigo23_amr_mtm,artigo24_amr_mtm,artigo27_amr_mtm,artigo28_amr_mtm,artigo29_amr_mtm,artigo30_amr_mtm,artigo35_amr_mtm}.
Among the studies that reported the pathogen, the most frequently modelled bacteria were \textit{Pseudomonas aeruginosa}, addressed in three studies \cite{artigo25_amr_mtm,artigo26_amr_mtm,artigo31_amr_mtm}, \textit{Escherichia coli}, examined in three studies \cite{artigo19_amr_mtm,artigo20_amr_mtm,artigo22_amr_mtm}, and \textit{Staphylococcus aureus}, modelled in three studies \cite{artigo07_amr_mtm,artigo33_amr_mtm,artigo34_amr_mtm}. Other pathogens included \textit{Acinetobacter baumannii}, in a single study \cite{artigo36_amr_mtm}, and a composite group comprising \textit{Clostridioides difficile}, \textit{Staphylococcus aureus}, \textit{Escherichia coli} and \textit{Klebsiella pneumoniae}, in a single study \cite{artigo32_amr_mtm}.

\subsection{Model construction and correlated parameters}

This section examines the model parameterisation, highlighting which biological complexities are commonly integrated and which are frequently omitted, potentially pointing to important gaps in the current modelling landscape.
In particular, it considers the representation of resistance acquisition mechanisms, the incorporation of PK/PD processes, and the inclusion of additional biological features such as fitness costs, host immune responses, spatial dispersion, and multiple resistance mechanisms. 

\begin{figure}[H]
  \centering
  \begin{subfigure}{0.48\textwidth}
     \includegraphics[scale=0.26]{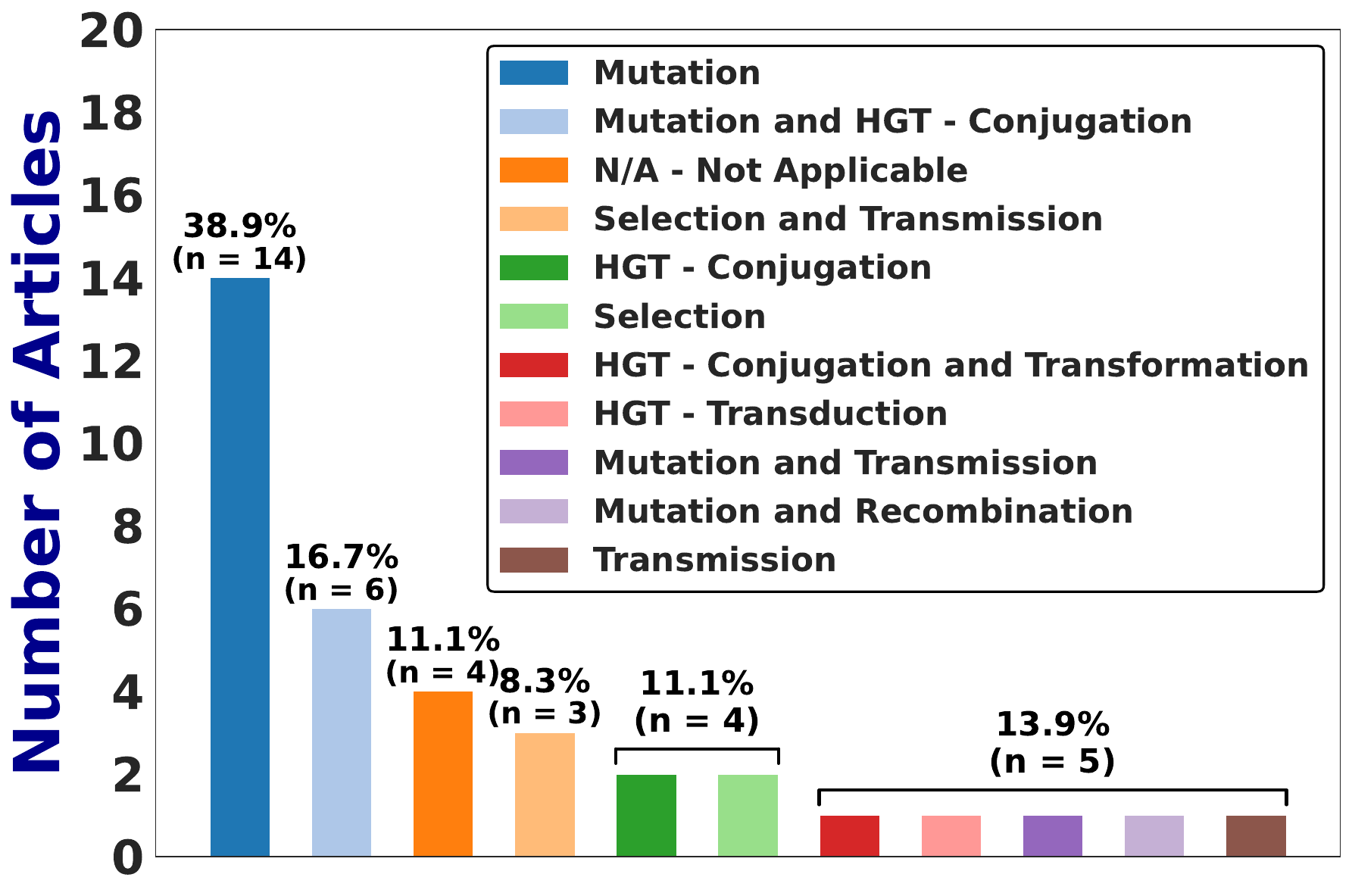}
     \caption{Mechanisms of acquired resistance}
     \label{fig:acquired_resistance}
  \end{subfigure}
  \quad
  \begin{subfigure}{0.48\textwidth}
     \includegraphics[scale=0.26]{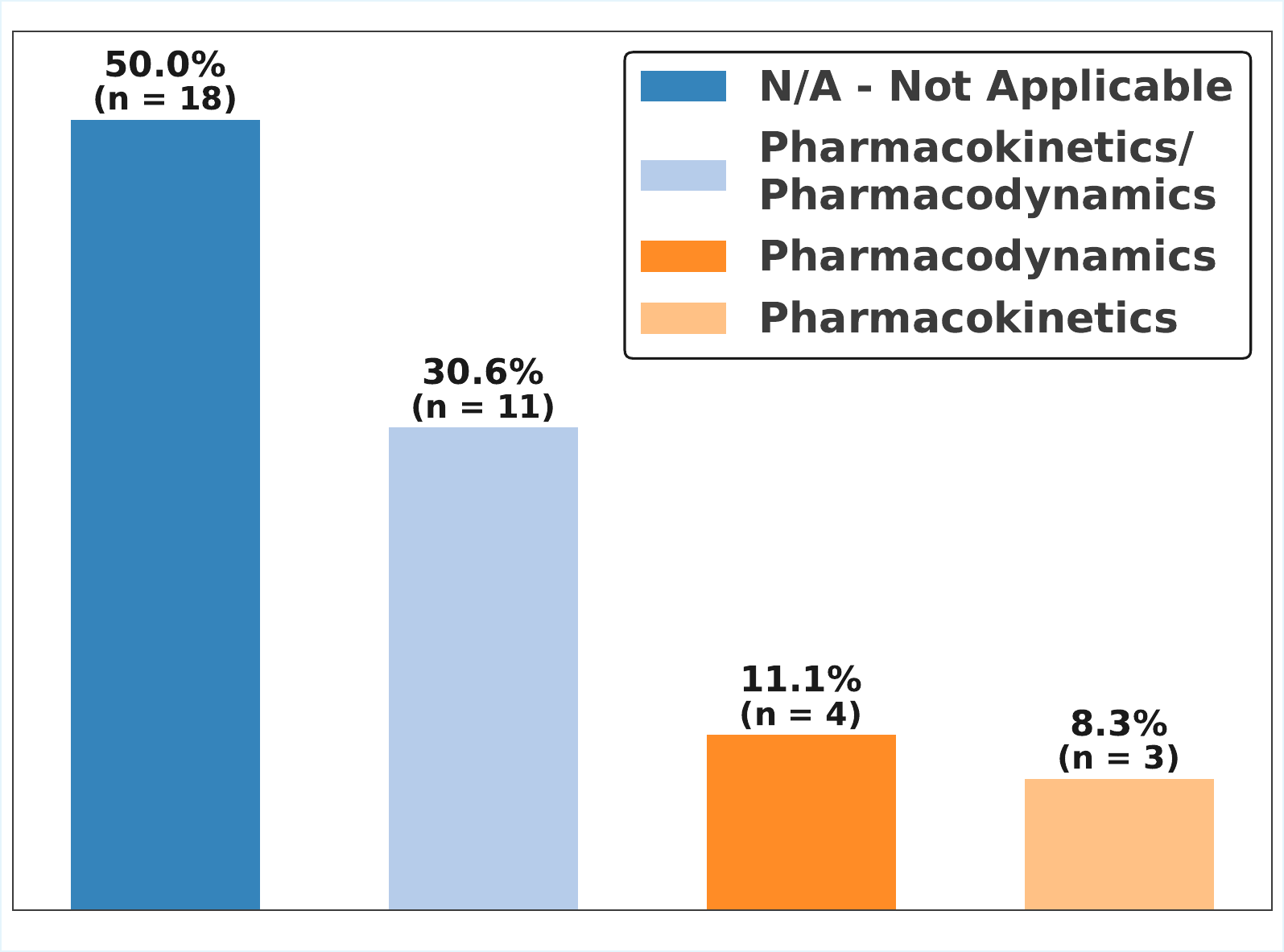}
     \caption{PK/PD modelling}
     \label{fig:pk_pd}
  \end{subfigure}

  \caption{\textbf{Mechanisms of AMR acquisition and integration of PK/PD modelling in the reviewed studies.} (a) Distribution of resistance acquisition mechanisms considered in the models, including mutation, horizontal gene transfer (HGT) via conjugation, transformation, or transduction, and models that incorporate multiple mechanisms or none. Mutation and HGT (especially via conjugation)  were the most frequently modelled mechanisms. (b) Overview of the use of PK/PD modelling approaches. Studies were classified according to whether they applied full PK/PD modelling, used only PK or PD components, or did not include these aspects at all. Nearly half of the studies did not incorporate any PK/PD modelling.}
  \label{fig:acquired_resistance_and_pk_pd}
\end{figure}

Figure \ref{fig:acquired_resistance_and_pk_pd} shows the results regarding resistance acquisition mechanisms and the use of PK/PD modelling in the thirty-six articles included in this review. 
In particular, regarding the mechanisms of resistance acquisition (Figure \ref{fig:acquired_resistance}), a total of $14$ studies $(38.9\%)$ considered mutation as the sole mechanism \cite{artigo05_amr_mtm,artigo07_amr_mtm,artigo08_amr_mtm,artigo10_amr_mtm,artigo15_amr_mtm,artigo16_amr_mtm,artigo19_amr_mtm,artigo22_amr_mtm,artigo24_amr_mtm,artigo25_amr_mtm,artigo26_amr_mtm,artigo28_amr_mtm,artigo30_amr_mtm,artigo35_amr_mtm}.
Six studies $(16.7\%)$ reported a combination of mutation and HGT via conjugation \cite{artigo04_amr_mtm,artigo11_amr_mtm,artigo13_amr_mtm,artigo14_amr_mtm,artigo18_amr_mtm,artigo33_amr_mtm}, whereas four studies $(16.7\%)$ did not specify any mechanism \cite{artigo01_amr_mtm,artigo02_amr_mtm,artigo17_amr_mtm,artigo20_amr_mtm}, three articles $(8.3\%)$ considered selection and transmission \cite{artigo03_amr_mtm,artigo09_amr_mtm,artigo31_amr_mtm}, two $(5.6\%)$ considered HGT via conjugation exclusively \cite{artigo12_amr_mtm,artigo32_amr_mtm}, two focused exclusively on selection \cite{artigo21_amr_mtm,artigo27_amr_mtm}. 
Other acquisition-related categories with one study $(2.8\%)$ each included HGT via conjugation and HGT via transformation \cite{artigo06_amr_mtm}, HGT via transduction \cite{artigo34_amr_mtm},  mutation and transmission \cite{artigo23_amr_mtm}, mutation combined with recombination \cite{artigo29_amr_mtm}, and transmission \cite{artigo36_amr_mtm}. Figure \ref{fig:pk_pd} summarizes how PK/PD modelling was incorporated in selected studies. Half of the studies $(18/36)$ did not apply PK/PD modelling (N/A -- not applicable) \cite{artigo01_amr_mtm,artigo03_amr_mtm,artigo04_amr_mtm,artigo09_amr_mtm,artigo10_amr_mtm,artigo11_amr_mtm,artigo13_amr_mtm,artigo14_amr_mtm,artigo17_amr_mtm,artigo18_amr_mtm,artigo21_amr_mtm,artigo24_amr_mtm,artigo27_amr_mtm,artigo29_amr_mtm,artigo30_amr_mtm,artigo31_amr_mtm,artigo32_amr_mtm,artigo36_amr_mtm}, $11$ studies $(30.6\%)$ combined PK/PD modelling in their analyses \cite{artigo02_amr_mtm,artigo05_amr_mtm,artigo06_amr_mtm,artigo07_amr_mtm,artigo12_amr_mtm,artigo15_amr_mtm,artigo16_amr_mtm,artigo20_amr_mtm,artigo25_amr_mtm,artigo26_amr_mtm,artigo34_amr_mtm}, four articles $(11.1\%)$ used only PD modelling \cite{artigo22_amr_mtm,artigo23_amr_mtm,artigo28_amr_mtm,artigo35_amr_mtm}, and three studies $(8.3\%)$ focused on PK modelling \cite{artigo08_amr_mtm,artigo19_amr_mtm,artigo33_amr_mtm}.

Mutation and HGT via conjugation are frequently modelled as the primary routes of resistance acquisition because they are specific, quantifiable, and relatively straightforward to incorporate into mathematical frameworks \cite{liu2024current,martinez2000mutation}. 
In contrast, HGT via transformation or HGT via transduction, while crucial in the spread of resistance, are more complex processes to model, involving factors such as exogenous DNA uptake (transformation) or viral infection (transduction) \cite{dimitriu2022evolution,maclean2015microbial}.
The lack of PK/PD data in many studies is partly due to the focus on the population dynamics of resistance, without explicit modelling of drug–host interactions. Furthermore, PK/PD modelling requires detailed experimental data, which are often unavailable or not the objective of the study, leading to a preference for simpler approaches \cite{rodriguez2021role,velkov2013pk}.

Table~\ref{tab:model_construction_and_parametrization} summarizes several important features concerning model construction and correlated parameters identified across the included studies. Most of the reviewed studies $(28/36; 77.8\%)$ reported a fitness cost associated with the resistant strain \cite{artigo04_amr_mtm,artigo05_amr_mtm,artigo06_amr_mtm,artigo07_amr_mtm,artigo08_amr_mtm,artigo09_amr_mtm,artigo10_amr_mtm,artigo11_amr_mtm,artigo12_amr_mtm,artigo13_amr_mtm,artigo14_amr_mtm,artigo15_amr_mtm,artigo16_amr_mtm,artigo19_amr_mtm,artigo20_amr_mtm,artigo23_amr_mtm,artigo24_amr_mtm,artigo25_amr_mtm,artigo26_amr_mtm,artigo27_amr_mtm,artigo28_amr_mtm,artigo29_amr_mtm,artigo30_amr_mtm,artigo31_amr_mtm,artigo32_amr_mtm,artigo33_amr_mtm,artigo34_amr_mtm,artigo35_amr_mtm}.
In contrast, other important features were largely absent, and only four studies considered spatial dispersion \cite{artigo20_amr_mtm,artigo21_amr_mtm,artigo27_amr_mtm,artigo29_amr_mtm}, and just eight included the host immune response \cite{artigo02_amr_mtm,artigo04_amr_mtm,artigo08_amr_mtm,artigo10_amr_mtm,artigo25_amr_mtm,artigo26_amr_mtm,artigo30_amr_mtm,artigo33_amr_mtm}.
The modelling of multiple resistance, although relevant to the biological reality of AMR, has been performed in only eight studies \cite{artigo07_amr_mtm,artigo13_amr_mtm,artigo15_amr_mtm,artigo18_amr_mtm,artigo21_amr_mtm,artigo29_amr_mtm,artigo32_amr_mtm,artigo34_amr_mtm}. 
Drug concentration, an essential element for capturing PK/PD processes, was explicitly incorporated as a dynamic variable in $18$ of the $36$ studies analysed \cite{artigo02_amr_mtm,artigo05_amr_mtm,artigo06_amr_mtm,artigo07_amr_mtm,artigo12_amr_mtm,artigo15_amr_mtm,artigo16_amr_mtm,artigo19_amr_mtm,artigo20_amr_mtm,artigo21_amr_mtm,artigo22_amr_mtm,artigo23_amr_mtm,artigo24_amr_mtm,artigo25_amr_mtm,artigo26_amr_mtm,artigo28_amr_mtm,artigo34_amr_mtm,artigo35_amr_mtm}. 
The remaining studies modelled treatment effects through simplified or empirical functions without including antibiotic concentration explicitly.

\begin{table}[htbp]
\centering
\caption{Summary of the presence or absence of model characteristics within the construction and correlated parameter categories across the 36 studies reviewed.}
\renewcommand{\arraystretch}{1.5}
\begin{tabular}{|>{\centering\arraybackslash}m{2.1cm}
                |>{\centering\arraybackslash}m{2.1cm}
                 >{\centering\arraybackslash}m{2.1cm}
                 >{\centering\arraybackslash}m{2.1cm}
                 >{\centering\arraybackslash}m{2.1cm}
                 >{\centering\arraybackslash}m{2.1cm}|}
\hline
\rowcolor{gray!25}\textbf{Presence} & \multicolumn{5}{c|}{\textbf{Biological Mechanism}} \\
\hline
\rowcolor{gray!25} & \textbf{Spatial dispersion} & \textbf{Fitness \ \ \ \ \ \ \  cost} & \textbf{Immune response} & \textbf{Multiple resistance} & \textbf{Drug concentration} \\
\hline
\textbf{Yes} & $4$ & $28$ & $8$ & $8$ & $18$  \\
\hline
\textbf{No} & $32$ & $8$ & $28$ & $28$ & $18$ \\
\hline
\end{tabular}
\label{tab:model_construction_and_parametrization}
\end{table}

These characteristics presented in Table \ref{tab:model_construction_and_parametrization} reveal important trends and limitations in the current practices of mathematical modelling of AMR. 
The frequent incorporation of fitness costs in models reflects their analytical tractability and solid foundation in the evolutionary theory of AMR \cite{melnyk2015fitness,vanacker2023fitness}. 
On the other hand, the host immune response is a central component in clearing infections and resistance, yet it is often neglected \cite{gjini2016integrating}.
A robust immune response can reduce the required duration of treatment and even alter the mutant selection range by limiting the drug concentration that favors partially resistant strains \cite{handel2009exploring}. 
Moreover, spatial dispersal has rarely been considered, despite being a key factor in the spread of resistance in heterogeneous environments, where host mobility, variability in treatment access, and unequal drug distribution create complex selective landscapes \cite{real2007spatial,chen2013temporal}. 
Additionally, the modelling of multiple resistance genes or mechanisms remains limited, despite its growing relevance in the face of emerging multidrug-resistant (MDR) and extensively drug-resistant (XDR) strains, as in the case of MDR/XDR tuberculosis \cite{basak2016multidrug,almutairy2024extensively}. 

\subsection{Model outputs and validation}

This subsection examines relevant aspects of model outputs and validation in the studies included in this review, such as the type of treatment used, which shapes the predicted selective pressures that drive resistance. 
Other important characteristics include the strategies adopted for model validation and outcome characterization, which demonstrate the robustness, applicability, and practical relevance of the modelling results. 
Analysing these aspects, we identify common practices, methodological preferences, and critical gaps (e.g., limited integration of economic impact) that prevent the models from comprehensively informing clinical and policy decision-making.

\begin{figure}[H]
  \centering
  \makebox[\linewidth]{
    \includegraphics[scale=0.3]{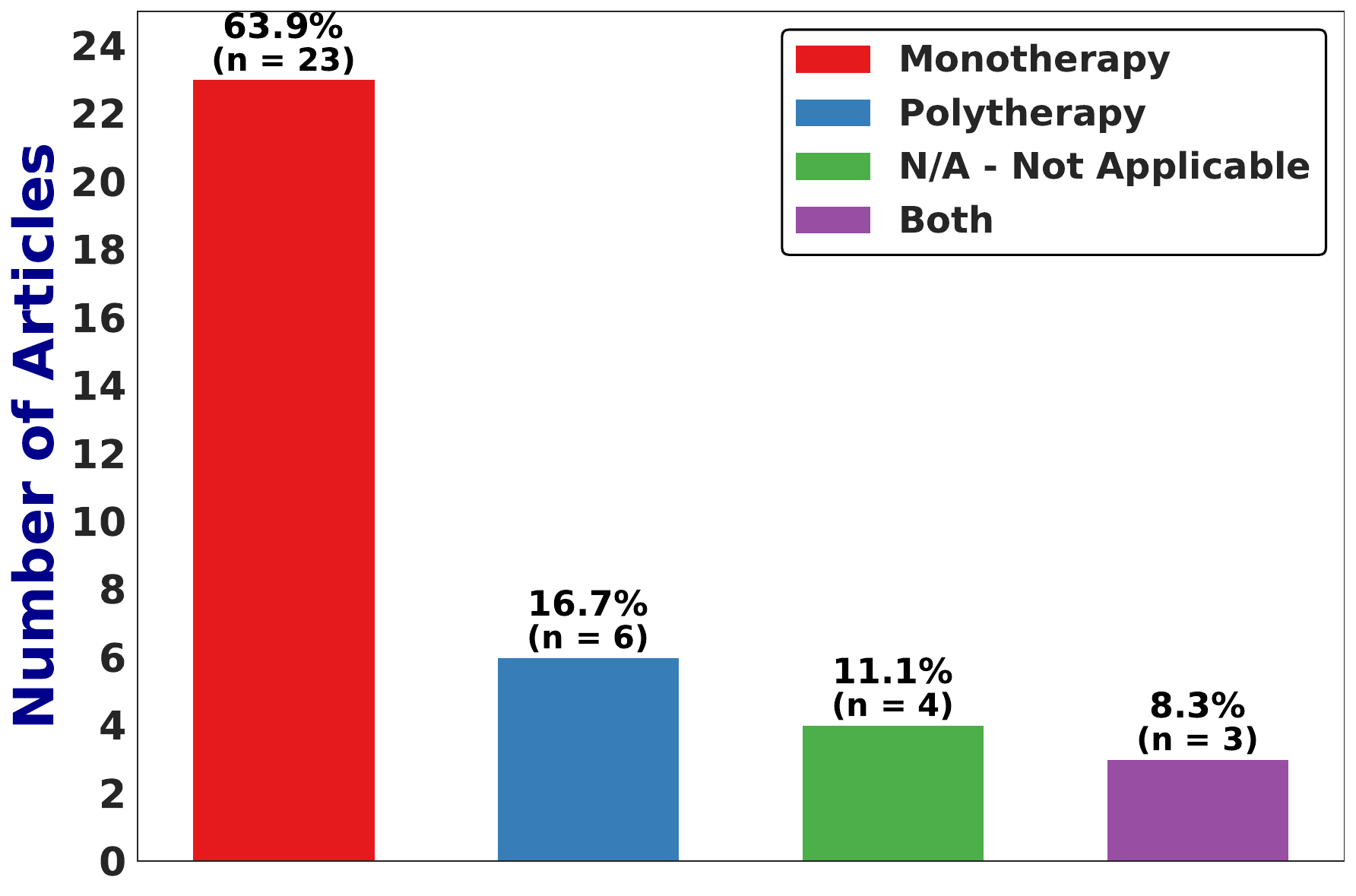}}
\caption{\textbf{Type of antimicrobial treatment modelled in the studies.} Among the 36 studies reviewed, monotherapy was the most common strategy used in $23$ studies. Six studies modelled polytherapy, four studies did not apply an antimicrobial treatment modelling (N/A -- not applicable), and three studies incorporated both approaches.}
  \label{fig:treatment_type}
\end{figure}

Figure \ref{fig:treatment_type} shows the types of antimicrobial treatment modelled in the included studies. Notably, the vast majority of studies $(23/ 36)$ adopted monotherapy \cite{artigo01_amr_mtm,artigo02_amr_mtm,artigo04_amr_mtm,artigo05_amr_mtm,artigo06_amr_mtm,artigo08_amr_mtm,artigo09_amr_mtm,artigo10_amr_mtm,artigo11_amr_mtm,artigo14_amr_mtm,artigo16_amr_mtm,artigo17_amr_mtm,artigo19_amr_mtm,artigo20_amr_mtm,artigo22_amr_mtm,artigo23_amr_mtm,artigo24_amr_mtm,artigo27_amr_mtm,artigo28_amr_mtm,artigo30_amr_mtm,artigo31_amr_mtm,artigo33_amr_mtm,artigo35_amr_mtm}, six studies $(16.7\%)$ used polytherapy \cite{artigo07_amr_mtm,artigo15_amr_mtm,artigo21_amr_mtm,artigo25_amr_mtm,artigo26_amr_mtm,artigo32_amr_mtm}, four studies did not apply an antimicrobial treatment modelling (N/A -- not applicable) \cite{artigo03_amr_mtm,artigo13_amr_mtm,artigo18_amr_mtm,artigo36_amr_mtm}, and three studies $(8.3\%)$ included both strategies (monotherapy and polytherapy) \cite{artigo12_amr_mtm,artigo29_amr_mtm,artigo34_amr_mtm}. Polytherapy in these models refers to the simultaneous use of two or more antimicrobial agents or the combination of an antimicrobial drug with an alternative therapeutic approach, such as bacteriophage therapy.

The frequent use of monotherapy in the included studies may reflect both practical and methodological factors. 
From a clinical point of view, monotherapy remains the standard treatment approach for many bacterial infections, particularly in contexts where resistance to first-line drugs is still limited or where surveillance systems are insufficient to guide combination therapy \cite{laxminarayan2013antibiotic,hsia2019use,goossens2005outpatient}. 
In terms of modelling approaches, monotherapy simplifies the representation of treatment dynamics and reduces the number of parameters required, facilitating model calibration and analysis \cite{spicknall2013modeling}. 
In addition, the lack of detailed empirical data on PK/PD interactions between drugs in combination regimens probably reduces the incorporation of polytherapy into mathematical models \cite{savoldi2021role}. 
These factors combined contribute to the increased adoption of monotherapy in mathematical models of AMR.

Table~\ref{tab:model_outputs_and_validation} summarizes relevant features concerning model outputs and validation identified across the review of included studies.
A sensitivity analysis was performed in twenty-seven studies \cite{artigo01_amr_mtm,artigo03_amr_mtm,artigo06_amr_mtm,artigo08_amr_mtm,artigo09_amr_mtm,artigo10_amr_mtm,artigo11_amr_mtm,artigo12_amr_mtm,artigo13_amr_mtm,artigo15_amr_mtm,artigo16_amr_mtm,artigo18_amr_mtm,artigo19_amr_mtm,artigo20_amr_mtm,artigo21_amr_mtm,artigo22_amr_mtm,artigo23_amr_mtm,artigo24_amr_mtm,artigo25_amr_mtm,artigo26_amr_mtm,artigo27_amr_mtm,artigo28_amr_mtm,artigo31_amr_mtm,artigo32_amr_mtm,artigo34_amr_mtm,artigo35_amr_mtm,artigo36_amr_mtm}, external validation is included in twenty-six studies \cite{artigo03_amr_mtm,artigo05_amr_mtm,artigo06_amr_mtm,artigo08_amr_mtm,artigo09_amr_mtm,artigo10_amr_mtm,artigo11_amr_mtm,artigo12_amr_mtm,artigo13_amr_mtm,artigo14_amr_mtm,artigo15_amr_mtm,artigo16_amr_mtm,artigo19_amr_mtm,artigo20_amr_mtm,artigo21_amr_mtm,artigo22_amr_mtm,artigo23_amr_mtm,artigo24_amr_mtm,artigo25_amr_mtm,artigo26_amr_mtm,artigo27_amr_mtm,artigo28_amr_mtm,artigo29_amr_mtm,artigo30_amr_mtm,artigo32_amr_mtm,artigo34_amr_mtm}, and almost all of the studies included evolution of resistance in their models, except for ten \cite{artigo01_amr_mtm,artigo02_amr_mtm,artigo04_amr_mtm,artigo07_amr_mtm,artigo17_amr_mtm,artigo18_amr_mtm,artigo31_amr_mtm,artigo33_amr_mtm,artigo35_amr_mtm,artigo36_amr_mtm}. 
On the other hand, although the economic impact of AMR is widely recognized, it is rarely incorporated into mathematical models, with only two studies explicitly including economic components in their formulations \cite{artigo09_amr_mtm,artigo18_amr_mtm}.
Moreover, thirty-four studies assessed the practical application of interventions \cite{artigo03_amr_mtm,artigo04_amr_mtm,artigo05_amr_mtm,artigo06_amr_mtm,artigo07_amr_mtm,artigo08_amr_mtm,artigo09_amr_mtm,artigo10_amr_mtm,artigo11_amr_mtm,artigo12_amr_mtm,artigo13_amr_mtm,artigo14_amr_mtm,artigo15_amr_mtm,artigo16_amr_mtm,artigo17_amr_mtm,artigo18_amr_mtm,artigo19_amr_mtm,artigo20_amr_mtm,artigo21_amr_mtm,artigo22_amr_mtm,artigo23_amr_mtm,artigo24_amr_mtm,artigo25_amr_mtm,artigo26_amr_mtm,artigo27_amr_mtm,artigo28_amr_mtm,artigo29_amr_mtm,artigo30_amr_mtm,artigo31_amr_mtm,artigo32_amr_mtm,artigo33_amr_mtm,artigo34_amr_mtm,artigo35_amr_mtm,artigo36_amr_mtm}.

\begin{table}[htbp]
\centering
\caption{Summary of the presence or absence of model characteristics within the outputs and validation categories across the 36 studies reviewed.}
\renewcommand{\arraystretch}{1.5}
\begin{tabular}{|>{\centering\arraybackslash}m{2.1cm}
                |>{\centering\arraybackslash}m{2.1cm}
                 >{\centering\arraybackslash}m{2.1cm}
                 >{\centering\arraybackslash}m{2.1cm}
                 >{\centering\arraybackslash}m{2.1cm}
                 >{\centering\arraybackslash}m{2.1cm}|}
\hline
\rowcolor{gray!25}\textbf{Presence} & \multicolumn{5}{c|}{\textbf{Output/Validation}} \\
\hline
\rowcolor{gray!25} & \textbf{Sensitivity analysis} & \textbf{External validation} & \textbf{Evolution of resistance} & \textbf{Economic impact} & \textbf{Practical applications} \\
\hline
\textbf{Yes} & $27$ & $26$ & $26$ & $2$ & $34$ \\
\hline
\textbf{No} & $9$ & $10$ & $10$ & $34$ & $2$ \\
\hline
\end{tabular}
\label{tab:model_outputs_and_validation}
\end{table}

All the information in Table \ref{tab:model_outputs_and_validation} is relevant when referring to the mathematical modelling of AMR. 
However, we highlight the scarcity of studies that incorporate the economic impact that AMR has on local and global economies.
It is estimated that, by 2050, global financial losses associated with AMR could range from $300$ billion to $1$ trillion dollars \cite{burki2018superbugs}, with accumulated losses of economic output in Organization for Economic Cooperation and Development (OECD) countries on the order of $20$ -- $35$ trillion \cite{review2014}. Moreover, the World Bank warns that AMR could increase poverty levels, with disproportionate effects on low- and middle-income countries (LMICs) \cite{bank2017drug}. 
In these countries, the higher costs of medical care are driven mainly by the need for prolonged treatment and additional hospitalizations. 
This reality is intensified by factors such as limited access to effective treatments, weak regulation, low awareness of the proper use of antibiotics, and inadequate dissemination of therapeutic guidelines \cite{bloom2017antimicrobial}.

\section{Discussion}
\label{sec12}
AMR represents one of the major contemporary global challenges to public health. 
Given the growing use of dynamic mathematical models to understand the mechanisms driving its emergence and spread, this scoping review synthesized evidence from $36$ studies published between $2019$ and $2024$ that employed mathematical modelling to investigate different aspects of AMR. 
From the results, we identified significant trends, prevalent methodological approaches, and essential research gaps. During the period analysed in the review, the number of annual publications varied from four to seven, demonstrating sustained and developing research activity in AMR.
The included studies originate from institutions in $20$ countries, with the United States, France, and the United Kingdom representing the countries with the most publications, highlighting both the influence of English-language search criteria and the concentration of publications in countries where greater research funding is allocated.

This geographical imbalance likely reflects structural inequities in global knowledge production rather than differences in scientific capacity, as evidenced by persistent disparities in research funding, publication networks, and access to data across regions \cite{king2004scientific,bhaumik2019diversity,worldbank2024rnd}.The predominance of studies included in this review being located in HICs can be attributed to several factors: (i) substantially greater investments in research and development (R\&D) in HICs (averaging $2.5\%$-$\$3\%$ of gross domestic product (GDP)) compared to most Global South nations ($<1\%$ of GDP), which enables advanced infrastructure for complex modelling research \cite{king2004scientific,worldbank2024rnd}; (ii) publication biases in high-impact journals, where editorial boards and peer review networks remain predominantly concentrated in North America and Europe \cite{bhaumik2019diversity}; and (iii) resource-dependent barriers, including limited access to specialized software, epidemiological data repositories, and computational power in low-research-funding environments \cite{agbeyangi2024advances,van2021unravelling}. 
Consequently, while our review includes valuable contributions from countries such as India and Egypt, research from these and other less-represented regions remains scarce. 
This imbalance not only limits the generalization of findings to diverse epidemiological contexts but also undermines the ability of mathematical models to generate policy-relevant insights for populations facing the highest burden of AMR, particularly in the Global South.

The predominance of bacterial pathogens in the reviewed models reflects their central role in the global burden of AMR \cite{naghavi2024global}. 
Bacteria are the primary targets of antimicrobial use in both human and animal health, and they account for the most pressing clinical challenges associated with drug resistance, including multidrug-resistant (MDR) and extensively drug-resistant (XDR) strains \cite{review2016, murray2022}.
Major global initiatives, such as the World Health Organization’s (WHO) Global Antimicrobial Resistance and Use Surveillance System (GLASS) \cite{world2022global}, the $2024$ WHO Bacterial Priority Pathogens List \cite{world2024bacterial}, and the United States Centers for Disease Control and Prevention’s (CDC) threat reports \cite{cdc2019threats}, prioritize bacterial species like \textit{E. coli}, \textit{S. aureus}, \textit{K. pneumoniae}, and \textit{P. aeruginosa} because of their high prevalence and resistance profiles. 
The modelling focus on these pathogens aligns with these priorities. In contrast, viral and fungal AMR have received less modelling attention, despite growing concerns over resistance in viruses such as HIV and influenza \cite{world2015global}, and in fungi such as \textit{C. auris}, which the CDC lists as an urgent threat \cite{lyman2023worsening}.
This highlights an important limitation in current modelling efforts, particularly the lack of frameworks that capture how different antibiotic classes interact with the host microbiota and influence pathogen dynamics, including non-bacterial agents.

The omission of the host immune response (included in only eight studies) and PK/PD modelling (absent in $50\%$ of studies) represents significant challenges for translating model predictions into clinical practice. 
Host immunity is central to infection clearance and influences the selection window for resistance. 
Moreover, the lack of PK/PD frameworks directly limits the capacity of models to assess drug-dosing strategies and optimize clinical outcomes, thereby reducing their usefulness in developing actionable therapeutic guidelines against MDR/XDR strains.

The scarcity of studies that integrate economic dimensions constitutes one of the most critical limitations for policy translation.
This omission represents a fundamental barrier, as without cost-effectiveness or macroeconomic analyses, models fail to provide the quantifiable evidence required by policymakers to justify substantial investments in AMR control and intervention programs, particularly in LMICs where the economic burden is relatively high.
Given the projected economic losses associated with AMR, including estimated annual reductions of up to $1\%$ in global GDP and even more severe impacts, between $5\%$ and $7\%$, in developing countries \cite{anderson2019averting,bank2017drug}, the integration of economic and epidemiological frameworks into future modelling efforts is essential to strengthen their policy relevance.
Furthermore, AMR-related effects are expected to extend to strategic sectors such as international trade and animal production, where increasing resistance to antibiotics used in livestock is projected to reduce the global supply of animal-derived food (e.g., meat, milk, and eggs) by approximately $11\%$ by $2050$, with the greatest repercussions in LMICs \cite{lekagul2019patterns,hao2014benefits,van2015global,bank2017drug}.

\subsection*{Key findings}
The findings of this scoping review reveal a consistent pattern in how mathematical modelling has been applied to study AMR over the past six years. 
The use of mathematical models to study AMR has remained active over the past six years, with a focus on deterministic models, particularly those based on ordinary differential equations (ODEs). 
Almost all of the included studies focused on bacterial pathogens, especially \textit{P. aeruginosa}, \textit{E. coli}, and \textit{S. aureus}, demonstrating global health priorities, given that bacteria are the main pathogens that cause AMR. 
In contrast, viral, fungal, and parasitic resistance remains neglected in mathematical modelling studies. 
Over half of the studies focused on human hosts, whereas animal and environmental contexts were rarely considered. 
Only one of the studies adopted a One Health perspective, employing a multiscale modelling approach. 
With respect to the mechanisms of resistance acquisition, mutation was the most frequently modelled pathway, followed by horizontal gene transfer (HGT), especially via conjugation, whereas transformation and transduction were rarely incorporated. Approximately half of the studies did not incorporate pharmacokinetic/pharmacodynamic (PK/PD) modelling. 
Few studies used PK/PD modelling simultaneously, while others have modelled only one of the components. Critical biological factors, such as host immune response and spatial dispersion, have rarely been explored and are underrepresented in the models, despite their epidemiological relevance. 
Therapeutic strategies were predominantly modelled as monotherapy, with combination therapies of two or more drugs rarely included in the models.
While external validation has commonly attempted, economic impact assessments and future scenario projections have rarely been included in model outputs. 
A significant geographic imbalance was observed, with most studies conducted by institutions in high-income countries (HICs). 
Contributions from the Global South remain limited, restricting the generalizability of the results to diverse epidemiological contexts.
These important findings show where mathematical modelling of AMR has focused and consolidated in recent years, highlighting its strengths and aspects that need further exploration, and guiding priorities for future model development and interdisciplinary collaboration.

\subsection*{Research gaps}

Our scoping review revealed several persistent methodological and conceptual gaps in current mathematical modelling studies of AMR. In the vast majority of the analysed articles, the authors chose to use deterministic methods mainly based on ODEs, ignoring stochasticity, rare events, and heterogeneity at the individual level which can influence the dynamics of AMR. Only one study incorporated explicit interactions among humans, animals, and the environment, which limits the overall capacity to represent feedback and AMR transmission routes across sectors according to the One Health framework. 
Similarly, viral, fungal, and parasitic AMR was practically not mentioned in the studies or mathematical models analysed, despite its growing clinical and public health importance.
Many models focus only on a single pathway of resistance acquisition (e.g., mutation or HGT—conjugation), ignoring the potential effects of combined mechanisms or other mechanisms such as HGT (transduction) and HGT (transformation). 
Few studies integrate simultaneously incorporated PK/PD frameworks, highlighting a significant research gap. 
Only a few studies have considered the economic impact that AMR can have on a country's GDP, disregarding the costs associated with increased morbidity and mortality, increased and longer hospital stays, higher treatment costs, and lost productivity.
In addition, spatial heterogeneity has rarely been considered, despite its potential role in shaping the emergence and spread of resistance. Addressing these gaps will be essential for developing more comprehensive, predictive, and policy-relevant frameworks for understanding and mitigating AMR.

\subsection*{Conclusion}

By mapping recent literature on the mathematical modelling of AMR, this scoping review revealed a developing field, but one still has significant methodological and conceptual discrepancies. We found a predominance of deterministic models based on ODEs, which, although useful for initial analysis, reveal a disconnect from the stochastic and heterogeneous nature of the evolution and spread of AMR. 
Similarly, the almost exclusive focus on bacterial pathogens and the mechanisms of mutation and HGT--conjugation overlooks essential components of the problem, such as AMR in viruses and fungi and the contribution of other gene transfer pathways.

Our findings show an underrepresentation of One Health models. The near absence of models that integrate human, animal, and environmental components prevents a comprehensive understanding of AMR transmission cycles. 
Furthermore, the rare incorporation of economic impact modelling in analyses limits the usefulness of models for policymakers who need to quantify the costs associated with AMR.

Therefore, we conclude that advances in mathematical modelling applied to AMR require conceptual and methodological shifts. Models must be able to capture the complexities associated with AMR, requiring the incorporation of stochastic and hybrid approaches, expanding their application to non-bacterial pathogens, and integrating the One Health perspective and socioeconomic variables. 
This conceptual shift is crucial for models to fully realize their potential as predictive tools and provide policymakers with the evidence needed to formulate more effective, realistic, and globally equitable intervention strategies.

\subsection*{Limitations and future directions}

This review is limited to articles published in English, which may exclude valuable contributions in other languages. Furthermore, although this scoping review focused on mapping trends and identifying gaps in the mathematical modelling of AMR, it did not include a critical assessment of methodological quality.
Additionally, no formal protocol was registered a priori in databases such as PROSPERO or OSF, which may reduce transparency and reproducibility, despite adherence to an internal protocol.
Moreover, only publications indexed in three databases (PubMed, Web of Science, and Scopus) were considered, and the specific Boolean search string used may have influenced the retrieval of studies, possibly omitting relevant works. 
The review period $(2019--2024)$ was selected to capture recent advances but may have excluded seminal older studies that remain methodologically relevant.

Another important limitation revealed by this review concerns the scarcity of mathematical representation of therapeutic strategies involving antibiotic combinations with distinct mechanisms of action. 
Although combination therapy and synergistic or antagonistic interactions are widely used in clinical practice to delay resistance, most models assume independent drug effects or restrict analyses to monotherapies. 
The incorporation of multi-drug interactions, including suppressive or competition-mediated effects, remains an open line of research with significant potential to improve predictions of resistance evolution.  
An additional aspect that none of the reviewed studies considered was the ecological impact of antimicrobial treatment on the host-associated microbiota. 
Antibiotic-induced dysbiosis can weaken colonization resistance, alter competitive dynamics, and indirectly promote the expansion or acquisition of resistant strains \cite{cusumano2025,shayista2025}. 
The absence of ecological interactions at the microbiota level in current mathematical models represents a significant gap, suggesting that future work should integrate this aspect to more realistically capture the emergence and persistence of AMR.

Future work should explore a broader inclusion of AMR modelling studies beyond the English-language literature and conduct comparative analyses between different modelling approaches (e.g., deterministic vs. stochastic models); develop more complex and realistic models that incorporate heterogeneous populations, multiple resistance genes, and treatments considering more than one type of drug; increase the use of stochastic, hybrid, or individual-based models to reflect real-world dynamics; incorporate components into the models that seek to assess the real economic impact that AMR has on a country; and encourage the development of models that integrate the human, animal, and environmental sectors in accordance with the One Health framework.
On the basis of the insights presented here, we also intend to develop a mathematical model of AMR in future work, aiming to address some of the gaps highlighted by this review.

\newpage

\subsubsection*{List of abbreviations}

\begin{itemize}
\item \textbf{ABMs} – Agent-based models
\item \textbf{ADME} – Absorption, distribution, metabolism, excretion
\item \textbf{AMR} – Antimicrobial resistance
\item \textbf{CDC} – Centers for disease control and prevention
\item \textbf{GDP} – Gross domestic product
\item \textbf{GLASS} – Global Antimicrobial Resistance and Use Surveillance System
\item \textbf{HGT} – Horizontal gene transfer
\item \textbf{HICs} – High-income countries
\item \textbf{HIV} – Human immunodeficiency virus
\item \textbf{ICU} – Intensive care unit
\item \textbf{IDEs} – Integro-differential equations
\item \textbf{JBI} – Joanna Briggs Institute
\item \textbf{LMICs} – Low- and middle-income countries
\item \textbf{LNA} – Linear noise approximation
\item \textbf{MDR} – Multidrug resistance
\item \textbf{MIC} – Minimum inhibitory concentration
\item \textbf{N/A} – Not applicable
\item \textbf{OECD} – Organization for Economic Cooperation and Development
\item \textbf{ODEs} – Ordinary differential equations
\item \textbf{OSF} – Open science framework
\item \textbf{PCC} – Population, concept, context
\item \textbf{PDEs} – Partial differential equations
\item \textbf{PK/PD} – Pharmacokinetic/Pharmacodynamic
\item \textbf{PRISMA-ScR} – Preferred reporting items for systematic reviews and meta-analyses extension for scoping reviews
\item \textbf{R\&D} – Research and development
\item \textbf{SDEs} – Stochastic differential equations
\item \textbf{WHO} – World health organization
\item \textbf{XDR} – Extensively drug-resistant
\end{itemize}

\subsubsection*{Pathogen abbreviations}

\begin{itemize}
  \item \textbf{\textit{A. baumannii}}: \textit{Acinetobacter baumannii}.
  \item \textbf{\textit{C. difficile}}: \textit{Clostridioides difficile}.
  \item \textbf{\textit{E. coli}}: \textit{Escherichia coli}.
  \item \textbf{\textit{K. pneumoniae}}: \textit{Klebsiella pneumoniae}.
  \item \textbf{\textit{P. aeruginosa}}: \textit{Pseudomonas aeruginosa}.
  \item \textbf{\textit{S. aureus}}: \textit{Staphylococcus aureus}.
  \item \textbf{\textit{S. pneumoniae}}: \textit{Streptococcus pneumoniae}.
\end{itemize}

\subsubsection*{Availability of data and materials}

The datasets generated and/or analysed during the current study are available in the GitHub repository: \href{https://github.com/felipeschardong/amr_modelling_scoping_review}{amr\_modelling\_scoping\_review}.

\subsubsection*{Competing interests}

The authors declare that they have no competing interests.

\subsubsection*{Authors' contributions}

FS contributed to the conception and design of the work, acquisition of data, analysis and interpretation of the data, creation of new software used in the study, and drafting of the manuscript. 
CLS contributed to the conception and design of the work, acquisition of data, interpretation of the data, revised the manuscript, and contributed to the writing. 
LMC contributed to the conception and design of the work, acquisition of data, interpretation of the data, revised the manuscript, and contributed to the writing. 
All authors read and approved the final manuscript and agree to be accountable for all aspects of the work.

\subsubsection*{Funding}

Part of this study was supported by the Coordination for the Improvement of Higher Education Personnel (CAPES) through a CAPES/PROSUP doctoral scholarship awarded to Felipe Schardong and by an additional scholarship provided by the Graduate Program of the School of Applied Mathematics at Getulio Vargas Foundation (FGV - EMAp).

\subsubsection*{Acknowledgements}

The authors would like to thank Juliana Silva Corrêa from the FGV-EAESP for enlightening discussions. 

\subsection*{Declarations}

Ethics, Consent to Participate, and Consent to Publish declarations: not applicable.

\newpage

\setcounter{tocdepth}{1} 

\begin{center}
\section*{Supplementary Material}
\end{center}
\addcontentsline{toc}{section}{Supplementary Material} 

\noindent
\textbf{Appendix A} \hspace{1em} Evaluation aspects of each study \dotfill \pageref{aspects_of_each_study} \\
\textbf{Appendix B} \hspace{1em} Code and data availability \dotfill \pageref{code_and_data} \\

\newpage

\appendix
\setcounter{table}{0}
\renewcommand{\thetable}{S\arabic{table}}
\renewcommand{\thefigure}{S\arabic{figure}}

\makeatletter
\renewcommand{\thesection}{\Alph{section}}
\renewcommand\@seccntformat[1]{Appendix~\csname the#1\endcsname\quad}
\makeatother

\section{Evaluation Aspects of Each Study}
\label{aspects_of_each_study}
To analyse the studies included in this scoping review, we developed a set of structured questions organized into three main categories: (i) Model Type and Context, (ii) Model Construction and Correlated Parameters, and (iii) Model Outputs and Validation. 
The purpose of these questions is to guide the extraction of relevant information about mathematical models of AMR, allowing consistent comparisons across studies. 
Each question was formulated to allow objective and comparable responses, ensuring transparency and consistency in the characterization of the evaluated models.

\begin{center}
 \textbf{MODEL: TYPE AND CONTEXT} 
\end{center}

\begin{enumerate}
    \item What is the nature of the model used in the study (deterministic, stochastic, agent-based, etc.)?
    
    \item What is the main mathematical structure of the model (e.g., ODEs, PDEs, etc.)?
    
    \item Which host species are considered (e.g., humans, animals, etc.)?
    
    \item At what spatial or contextual scale is the model applied (e.g., individual, hospital, community, etc)?
    
    \item Which classes of pathogens are modelled (e.g., bacteria, viruses, fungi, parasites)?
    
    \item What is the specific pathogen considered in the model, when applicable (e.g., \textit{Mycobacterium tuberculosis}, \textit{Staphylococcus aureus})?
\end{enumerate}

\begin{center}
    \textbf{MODEL: CONSTRUCTION AND CORRELATED PARAMETERS}
\end{center}

\begin{enumerate}
    \item Does the model include explicit spatial components (e.g., contact networks, dispersal between regions)?  

    \item Does the model incorporate fitness costs (adaptive advantages/disadvantages) for resistant strains?  

    \item Does the model include host immune response dynamics?  

    \item Which mechanisms of resistance acquisition are considered (e.g., mutation, horizontal gene transfer - HGT, etc.)?  

    \item Does the model address resistance to multiple antimicrobials simultaneously (multidrug resistance)?  

    \item Does the model incorporate pharmacokinetic/pharmacodynamic (PK/PD) parameters to describe antimicrobial action?  
\end{enumerate}

\begin{center}
    \textbf{MODEL: OUTPUTS AND VALIDATION}
\end{center}

\begin{enumerate}
    \item Was a sensitivity analysis performed to assess the influence of model parameters?  

    \item Are model results validated with empirical data or compared to existing models?  

    \item Does the model capture the temporal dynamics of resistance under prolonged antimicrobial exposure?  

    \item Does the model use antimicrobial efficacy parameters (e.g., Minimum Inhibitory Concentration - MIC)?  

    \item Does the model evaluate the cost-effectiveness or economic impacts of the analysed interventions?  

    \item Does the model consider monotherapy, combination therapy, or both?  

    \item Can the model results be used in practical applications (e.g., scenario forecasting, decision support, intervention planning)?
\end{enumerate}

\section{Code and data availability}
\label{code_and_data}

The code and data used in this study are available in the GitHub repository: 
\href{https://github.com/felipeschardong/amr_modelling_scoping_review}{amr\_modelling\_scoping\_review}.




\newpage

\bibliography{sn-bibliography}

\end{document}